# Comparison of mass renormalization schemes for simple model systems

E.V. Stefanovich[1], R.E. Wagner[2], Q. Su[2] and R. Grobe[2]

(1) 2255 Showers Drive, Unit 153, Mountain View, CA 94040 USA

(2) Intense Laser Physics Theory Unit and Department of Physics,
Illinois State University, Normal, IL 61790-4560 USA

We discuss an alternative method to mass renormalize a quantum field Hamiltonian based on a requirement that the vacuum and single-particle sectors are not self-scattering. We illustrate the feasibility of this method for the concrete example of bosonic quadratic and also quartic interactions. The results are compared with those obtained by the standard renormalization technique based on a spectral analysis of the usual field-based propagators. We also discuss a potential method for renormalizing a quantum field theory in a non-perturbative way using computational methods.

# 1. Introduction

The ultimate goal of our work is to develop a methodology that would permit us to study the quantum field theoretical interaction of electrons, positrons and photons with full space-time resolution. However, there are numerous conceptual and also computational challenges that need to be addressed first. Presently, all experimentally verified predictions of quantum electrodynamics such as transition rates and cross-sections are based on the matrix elements of the scattering operator [1,2]. This approach relates the properties of asymptotic incoming and outgoing scattering states and it is difficult to see how it could be used to also provide temporal and spatial information about the dynamics inside the interaction region. The later could be read off the time-dependent field theoretical state $|\Psi(t)\rangle$. While the QED Hamiltonian would describe the entire time evolution of $|\Psi(t)\rangle$, its energy eigenvalues and eigenvectors cannot be used immediately to construct the time evolution. This problem exists on top of the conceptual difficulty of converting the information contained in $|\Psi(t)\rangle$ to meaningful physical quantities that should at least in principle be observable. In order to overcome the singularities of the original Hamiltonian, it needs to be renormalized first [3,4]. This step can usually be performed perturbatively, but there are serious theoretical questions about the convergence of the complete perturbative series even after each term is renormalized, although this issue does not prevent practical application of perturbative methods in QED.

Many textbooks discuss renormalization techniques in an abstract way and concrete examples that apply these techniques for specific Hamiltonian systems are rare [5-9]. If the system is renormalizable, then the free (bare) parameters (such as the mass, charge or coupling constant) in the Hamiltonian can be adjusted in order to modify the spectrum of the Hamiltonian to fit experimental measurements of the physical parameters. This can be expressed equivalently by the construction of appropriate additional interactions (counterterms) to the Hamiltonian, where the coefficients of the counterterms are taken to be the difference between the bare and physical values of the parameters.

The motivation for this work is three-fold. While the general abstract principles of renormalizing a system on a Hamiltonian level are standard in most texts on quantum field theory, there is only a very limited number of specific examples how these abstract concepts are applied to concrete systems. Secondly, we introduce a new way of constructing counterterms so that the S-matrix operator satisfies a no-self-scattering condition for the vacuum and single particle states.

Finally, we introduce a new, albeit purely computational method to numerically renormalize a Hamiltonian. This method is non-perturbative and could be used as a yardstick to determine the accuracy of perturbative methods.

The work is structured as follows. In section 2 we describe the two model Hamiltonians used in this work and comment critically on the relationship between manifest co-variance and relativistic invariance as defined by the Poincare relationships. In section 3 we first briefly review the usual renormalization technique and introduce a new alternative technique that is based on imposed self-scattering conditions. It requires the determination some of the S-matrix elements and the associated self-scattering conditions for some free energy eigenstates lead to the addition of counterterms to the original Hamiltonian. The remainder of the work is devoted to providing concrete examples for this new renormalization method illustrating the commonalities and also differences to the usual approach. In section 4 we compare both approaches for the quadratic interaction, which while seemingly trivial it has exact solutions, which makes it an ideal illustration of the two approaches to renormalization. In section 5 we discuss the more general quartic interactions and discuss the new role of single- and two-particle masses for renormalization. Here we compare estimates based on perturbation theory and the usual Feynman techniqe with numerical results obtained from a direct diagonalization of the Hamiltonian in a truncated Hilbert space. We also apply the alternative S-matrix based approach to this system. In section 6 we provide a brief summary and an extended critical discussion of the results, its relevance for QED, open questions and future work.

**2. The model Hamiltonians and their Poincare invariance**

To construct a quantum field theory, one can equivalently take either the field operators or the particle creation and annihilation operators as the more fundamental operators. Below we adopt the particle-based approach, similar to what is used in Weinberg's book [10], which seems to us to be the closest to an ideal axiomatic-like method in the sense that it relies on the smallest number of heuristic assumptions. We restrict our discussion here to a scalar boson in one spatial direction, but generalizations to bosonic and fermionic systems such as QED in higher dimensions should be straightforward. The unperturbed, free Hamiltonian is given by

$$H_0 = \int dp \, \omega(p;m) \, \hat{a}^\dagger(p;m) \, \hat{a}(p;m) \tag{2.1}$$

where the usual bosonic annihilation and creation operators fulfill the commutator relation $[\hat{a}(p_1;m), \hat{a}(p_2;m)^\dagger]_- = \delta(p_1-p_2)$, the energy is given by $\omega(p;m) \equiv \sqrt{[m^2c^4+c^2p^2]}$, and m is the bare mass of the boson. Here and below we use $\hbar=1$ a.u. We denote with $|0;m\rangle$ and $\hat{a}^\dagger(p;m)|0;m\rangle$ the bare eigenvectors of $H_0$ with energies 0 and $\omega(p;m)$. This Hamiltonian $H_0$ is relativistically invariant, and we show in the Appendix A that using the operators $\hat{a}^\dagger(p;m)$ and $\hat{a}(p;m)$ one can explicitly construct the two translation operators for the position and velocity P and $K_0$ such that the three Poincare relationships $[H_0,P] = 0$, $[K_0,P] = -i H_0/c^2$ and $[K_0,H_0] = -i P$ are fullfilled.

Weinberg's method of constructing an interaction that leads to a Lorentz invariant scattering matrix requires two steps. First, he constructs the field operator from the particle creation and annihilation operators (in the $H_0$-based interaction picture)

$$\hat{\varphi}(z,t) \equiv (4\pi)^{-1/2} c \int dp\, \omega(p)^{-1/2} [\hat{a}(p) \exp(ipz-i\omega t) + \hat{a}^\dagger(p) \exp(-ipz+i\omega t)] \quad (2.2)$$

It was constructed in such a way that it transforms under the Lorentz transformation as a scalar, $\exp[-iK_0c\theta]\, \hat{\varphi}(z,t)\, \exp[iK_0c\theta] = \hat{\varphi}(L_{-\theta}[z,t])$, where the linear operation $L_{-\theta}[z,t]$ describes the inverse of the usual Lorentz transformation defined as $L_\theta(a,b) \equiv (a\,\mathrm{Cosh}\theta - b/c\,\mathrm{Sinh}\theta,\, b\,\mathrm{Cosh}\theta - a/c\,\mathrm{Sinh}\theta)$ with the rapidity parameter $\theta \equiv \tanh^{-1}(v/c)$. Manifest covariance means that applying the boost operator to the field can be performed by simply replacing the parameters z and t by the Lorentz transformed ones. In other words, the boosted field has the same functional form as the unboosted one, except it is now a function of $z'$ and $t'$ instead of z,t.

As a second step we construct a general manifestly co-variant Hamiltonian density in the interaction picture as a polynomial of this field and its derivatives with respect to the parameters z and t. More specifically, in this work we will examine the quadratic and quartic interactions $H = H_0 + \lambda \int dz\, \hat{\varphi}(z)^N$ for N=2 and 4. Even though we have not been able to prove its Poincare invariance, we will also discuss below the normal ordered form $V_{N4} \equiv N[\int dz\, \hat{\varphi}(z)^4]$, where N denotes normal ordering of creation and annihilation operators.

## 3. General approaches to renormalization

It turns out that due to the quadratic or quartic interaction, the parameters m and $\lambda$ are different from the observable mass and the real coupling (or charge) of the physical particles. These two observables must be correctly matched to the experimental values for a theory that contains interactions. In order to repair the theory, one has to first find how the parameters m and $\lambda$ are related to the physical observables. To make a better comparison, we first briefly review the basic ideas for the usual field- and propagator-based approach to renormalization and then introduce in Sec. 3.2 our proposed scattering-matrix based approach.

### 3.1 Lagrangian- and field-based approach to renormalization

As one of the key equations in quantum field theory, the LSZ reduction formula relates the scattering amplitude between the asymptotic initial and final states to the time-ordered vacuum expecation value of products of the interacting field operator [11,12]. Its validity requires that the vacuum expectation value of the field vanishes $\langle \text{VAC}|\hat{\varphi}(0)|\text{VAC}\rangle = 0$, and that the probability amplitude for the field to create or annihilate a one-particle energy eigenstate of H with momentum P takes the same value as it does in a free-theory, $\langle P|\hat{\varphi}(0)|\text{VAC}\rangle = \langle p|\hat{\varphi}_{\text{free}}(0)|0\rangle$. Here $|\text{VAC}\rangle$ is defined as the vacuum state of the interacting Hamiltonian, which is also the state of lowest energy. If we are to use the Feynman rules to calculate the S-matrix elements, we should first renormalize the fields to satisfy these two requirements [13]. The first condition can be accomplished by adding a constant to the field $\hat{\varphi} \rightarrow \hat{\varphi} + v$ for some vacuum expectation value v, and the second condition is accomplished by a multiplicative constant, $\hat{\varphi} \rightarrow \sqrt{Z}\hat{\varphi}$, as discussed below.

As a result, the renormalized field is one whose associated Green's function has the same behavior near its pole as for a free field, and the renormalized mass (M) is defined by the position of the pole. In this traditional method it is customary to use the Lagrangian framework and try

$$L = \tfrac{1}{2}(\partial_{ct}\hat{\varphi}_0)^2 - \tfrac{1}{2}(\partial_z\hat{\varphi}_0)^2 - \tfrac{1}{2}m^2c^2\hat{\varphi}_0^2 - \lambda\hat{\varphi}_0^N \qquad (3.1)$$

Here it is important to note that the free parameters m and $\lambda$ are not the physical mass or coupling.

In order to re-express the Lagrangian in terms of the physical quantities, we add and subtract terms from the Lagrangian such that we can write the *same* Lagrangian density of Eq. (3.1) as:

$$L = (1/2)(\partial_{ct}\hat{\varphi}_r)^2 - (1/2)(\partial_z\hat{\varphi}_r)^2 - (1/2) M^2c^2\hat{\varphi}_r^2 - \lambda\hat{\varphi}_r^N \text{ Counter}(m,\lambda) \quad (3.2)$$

Where the form of the counterterm Counter(m,λ) is

$$\text{Counter}(m,\lambda) = (1/2)\{(\partial_{ct}\hat{\varphi}_0)^2 - (\partial_{ct}\hat{\varphi}_r)^2\}$$
$$- (1/2)\{(\partial_z\hat{\varphi}_0)^2 - (\partial_z\hat{\varphi}_r)^2\} - (1/2)\{m^2c^2(\hat{\varphi}_0)^2 - M^2c^2(\hat{\varphi}_r)^2\} - \lambda\hat{\varphi}_0^N \quad (3.3)$$

The counterterm contains both $\hat{\varphi}_0$ and $\hat{\varphi}_r$. To simplify the term Counter(m,λ) an additional and possibly restrictive condition is used, namely, it is assumed that $\hat{\varphi}_0$ and $\hat{\varphi}_r$ differ by a constant of proportionality, $\hat{\varphi}_0 = Z(m,\lambda)^{1/2}\hat{\varphi}_r$. This condition is imposed in order to satify the relation described above that the field operator creates a single particle state, $\langle p|\hat{\varphi}_r(0)|0\rangle = \langle P|\hat{\varphi}(0)|VAC\rangle$. Usually it is then assumed that instead of merely finding m and λ we now have to find the values of three parameters. The bonus, however, is that we can now eliminate $\hat{\varphi}_0$ from our counterterm and everything depends only on one field. The counterterms then take the form Counter(m,λ) = $(Z-1)(1/2)\{(\partial_{ct}\hat{\varphi}_r)^2 - (\partial_z\hat{\varphi}_r)^2\} - (1/2)\{m^2c^2 Z - M^2c^2\}(\hat{\varphi}_r)^2\} - \lambda(Z^{1/2}\hat{\varphi}_r)^N$. So if we define A≡Z–1, B≡{$m^2c^2 Z - M^2c^2$} and C≡$\lambda Z^{N/2}$, we arrive at the counterterms 1/2 A{$(\partial_{ct}\hat{\varphi}_r)^2 -(\partial_z\hat{\varphi}_r)^2$} – 1/2 B$(\hat{\varphi}_r)^2$ – C $\hat{\varphi}_r^N$.

In other words, the counterterm is now a function of three unknown parameters. In order to compute the values we require certain properties of the two- and four- point correlation functions. If we define a new field operator $\hat{\varphi}(z,t) \equiv \exp[-iHt]\hat{\varphi}_r(z)\exp[iHt]$ and use the lowest energetic energy eigenstate of the Hamiltonian $H|VAC\rangle = E_{VAC}|VAC\rangle$, one can construct the two-point correlation function $\langle VAC|T\hat{\varphi}(z_1,t_1)\hat{\varphi}(z_2,t_2)|VAC\rangle$, where T is the time ordering operator. This function is a central quantity in the field-based approach to relativistic quantum mechanics advocated in most textbooks, and its Fourier transform is the propagator in 4-momentum space. Its pole is located at exactly the energy $Mc^2$ and the residue must be unity, consistent with the relation $\langle p|\hat{\varphi}_r(0)|0\rangle = \langle P|\hat{\Phi}(0)|VAC\rangle$ discussed above. The numerical value of the coupling $\lambda$ is to be determined from the four-point correlation function for a $\hat{\varphi}^4$ theory.

The usual method to relate the correct vacuum state $|VAC\rangle$ to the bare state $|0\rangle$ invokes the adiabatic theorem [14] according to which a time-evolved state remains an eigenstate if the interaction is only gradually turned on, $V_\eta(t) \equiv V \exp[-\eta|t|]$. We then obtain $|VAC\rangle = \lim_{\eta \to 0} \exp[-i \int d\tau\, H(\tau)] |vac\rangle$, where the lower integration limit is $-\infty$.

## 3.2 Alternative approach based on zero-scattering conditions

In contrast to the spectrally based method where the eigenenergies of the bare Hamiltonian are computed in order to invert the relationships between the bare parameters and the desired physical parameters, in the S-matrix based method proposed here we equate the bare parameters from the very beginning with the physical ones [15-17]. We want to assign some physical reality to our original states $|vac\rangle$ and $\hat{a}^\dagger(p_1)|vac\rangle$. As a result, the usual annihilation and creation operators are associated with the corresponding physical states. In this method, the interaction terms need to be constructed in such a way that the no-self-scattering property of the states $|0\rangle$ and $\hat{a}^\dagger(q)|0\rangle$ is guaranteed. Here, no-self-scattering means that the S-matrix carries these states into themselves, $\langle 0|S|0\rangle = 1$ and $\langle q'|S|q\rangle = \delta(q-q')$, where the state $|q\rangle$ is given by $|q\rangle \equiv \hat{a}^\dagger(q)|0\rangle$.

In this approach, the central quantity is the scattering-operator, defined as $S = \lim_{(t \to \infty,\, t_0 \to -\infty)}$ where

$$S(t,t_0) = 1 - i \int^t dt_1\, V(t_1) - \int^t dt_1\, V(t_1) \int^{t_1} dt_2\, V(t_2) + \ldots\} \qquad (3.4)$$

where $V(t) \equiv \exp[iH(t-t_0)]\, V \exp[-iH(t-t_0)]$ and the lower integration limit is $t_0$. Here we are still in the Schrödinger picture and the states are time-independent. It turns out that this scattering operator allows one to effectively split off the interaction-based portion from the full time evolution operator, $\exp[i H(t-t_0)] = \exp[-i H_0(t-t_0)]\, S(t,t_0)$. This allows a consecutive propagation for any state first by S(t,t₀), which can describe transitions between bare eigenstates of $H_0$, and then consecutively the free evolution associated solely with $H_0$.

A physical S-operator must be cluster separable so that S-matrix elements representing separate processes at long distances from each other do not contain any dependencies between spatially separated processes. Furthermore, it is assumed that the interaction is adiabatically turned on and off at t= ±∞. These additional assumptions allow one to assume that V(±∞)=0, leading to the vanishing commutator $[H_0,S]=0$ and the Poincare invariance of S.

The goal of the renormalization technique presented in this section is to modify the Hamiltonian $H=H_0+V$ by adding appropriate additional counterterms, such that the original bare vacuum state |vac⟩ as well as the bare single particle state $\hat{a}^\dagger(p)$|vac⟩ are kept unchanged by the scattering operator. The resulting (renormalized) Hamiltonian will correspond to a scattering operator, which fulfills the no-self-scattering conditions

$$S |0\rangle = |0\rangle \tag{3.5a}$$

$$S \hat{a}^\dagger(p_1) |0\rangle = \hat{a}^\dagger(p_1) |0\rangle \tag{3.5b}$$

For the theory to be renormalizable, we must be able to satisfy the no-self-scattering requirements of Eqs. (3.5) without needing to introduce new counterterms which are not part of the original bare Hamiltonian.

## 4. First example: The quadratic interaction

In order to illustrate the S-matrix based method proposed here and compare its predictions with that of the standard field-based approach, we use the simplest quantum field theory that predicts the undesirable self-scattering of the bare vacuum as well as the bare single-particle sector. The Hamiltonian is given by

$$H(m,\lambda) = \int dp\, \omega(p;m)\, \hat{a}^\dagger(p;m)\, \hat{a}(p;m) + \lambda \int dz\, \left(\hat{\varphi}(z)\right)^2 \tag{4.1}$$

This Hamiltonian is simple enough that one could immediately compute its energy spectrum exactly. The vacuum's expected energy is given by $E_{vac} = \delta(0) \int dp \, \{[\omega(p,m_e) - \omega(p;m)]^2 + 2\lambda/c^2\}/[4\omega(p;m_e)]$, where the effective mass is $m_e \equiv [1 + 2\lambda/(m^2 c^2)]^{1/2} m$. While the first term $\int dp \, [\omega(p,m_e) - \omega(p;m)]^2$ is finite for any $\lambda$, (which is contained in $m_e$), the integral over the second term proportional to $\int dp / [\omega(p;m_e)]$ diverges logarithmically. For a comparison which will be needed below we should mention the perturbative expansion of the energy $E_{vac} = \delta(0) \lambda/(2c^2) \int dp/\omega(p;m) + O(\lambda^2)$. The exact single particle energy is given by $E_{vac} + \omega(p,m_e)$. As we are focused here on illustrating the general method, for which the exact spectrum is usually unknown, we will not use this knowledge about $E_{vac}$, $\omega(p,m_e)$ and $m_e$ below.

## 4.1 Spectral approach for the single-particle mass

In this approach, we try to relate the (free) bare parameters m and $\lambda$ to the known value of the physical mass, denoted by M. There are several methods to do so, ranging from Feynman diagrams to numerical diagonalization to simple perturbation theory in the bare parameters. At this level, the standard renormalization treatment stops and uses these computed bare parameters for S-matrix calculations. However, in order to compare this traditional approach with our alternative approach in which a new Hamiltonian (called $H_{phys}$) is obtained, as a second step we need to rewrite the same Hamiltonian in terms of the more easily interpretable operators $\hat{a}(p;M)$. This rewriting will allow us to identify a new interaction in the original Hamiltonian. Note that the Hamiltonian is expressed in terms of the operators $\hat{a}(p;m)$ that create a particle with bare mass m. We denoted with $|0;m\rangle$ the bare vacuum and with $|q;m\rangle \equiv \hat{a}^\dagger(q;m) |0;m\rangle$ the bare single particle state with momentum q. Using simple first order perturbation theory in $\lambda$, we find $\langle 0;m|H|0;m\rangle = \lambda (c^2/2) \delta(0) \int dp \, \omega(p)^{-1}$ and $\langle q;m|H|q;m\rangle = \lambda (c^2/2) \delta(0) \int dp \, \omega(p)^{-1} + \omega(p;m) + \lambda (c^2/2) \omega(p)^{-1}$. It should be noted that the correction to the vacuum energy is infinite, associated with the logarithmic divergence of the integral $\int dp \, \omega(p)^{-1}$. The second infinite factor, $\delta(0)$, becomes $L/(2\pi)$ in a finite box of total length L. This factor is expected for a state that is translationally invariant and represents a vacuum energy density.

We subtract the vacuum energy $\langle 0;m|H|0;m\rangle$ first and then define the $O(\lambda^1)$ effective mass as $Mc^2 = \langle q;m|H|q;m\rangle|_{q=0} = [\omega(q;m) + \lambda\, c^2\, \omega(q)^{-1}]\,|_{q=0} = m\,c^2 + \lambda\, m^{-1}$. This equation determines the correct value of the bare mass, which we denote by $m_1$. If we invert this solution up to $O(\lambda^1)$, we find that the bare mass has to take the specific value $m_1$ given by

$$m_1 = M - \lambda/(Mc^2) + O(\lambda^2) \qquad (4.2)$$

Now as a second step we rewrite the same Hamiltonian $H(m_1,\lambda)$ in terms of more meaningful operators that create a particle of mass M. In general, the corresponding operators that create or annihilate a particle with either mass $m_1$ or M are related by the Bogoliubov-Valatin transformation [18]:

$$\hat{a}(p;m_1) = (\Omega+\omega_1)(4\Omega\omega_1)^{-1/2}\,\hat{a}(p;M) + (\omega_1-\Omega)(4\omega_1\Omega)^{-1/2}\,\hat{a}^\dagger(-p;M) \qquad (4.3)$$

with the energies $\omega_1(p) \equiv [m_1^2 c^4 + c^2 p^2]^{1/2}$ and $\Omega(p) \equiv [M^2 c^4 + c^2 p^2]^{1/2}$. If we apply this equation to our situation for mass $m_1 = M - \lambda/(Mc^2)$ and stay within $O(\lambda^1)$, we find $\hat{a}^\dagger(p; m_1) = \hat{a}^\dagger(p;M) - \lambda c^2/(2\Omega^2)\,\hat{a}(-p;M)$. If we then replace the bare operator $\hat{a}(p;m_1)$ by the more physical ones $\hat{a}(p;M)$, which we denote by $\hat{A}(p) \equiv \hat{a}(p;M)$, the original Hamiltonian becomes

$$H(m_1,\lambda) = \int dp\, \omega_1\, \{\hat{A}^\dagger(p)\,\hat{A}(p) - \lambda\, c^2/(2\Omega(p)^2)\,[\hat{A}^\dagger(p)\,\hat{A}^\dagger(-p) + \hat{A}(p)\hat{A}(-p)]\}$$
$$+ \lambda\, c^2/2\, \int dp\, \Omega(p)^{-1}\,[A^\dagger(p)A^\dagger(-p) + A^\dagger(p)A(p) + A(p)A^\dagger(p) + A(p)A(-p)] \qquad (4.4)$$

Note that we have expanded the interaction $\lambda \int dz\, (\hat{\varphi}(z))^2$ immediately in terms of the $A(p)$ and $A^\dagger(p)$. The common prefactor before the term $\hat{A}^\dagger(p)\,\hat{A}(p)$ is given by $\omega_1 + \lambda\, c^2 \Omega^{-1}$, which is then equivalent to $\Omega + O(\lambda^2)$. Furthermore, if we also use $-\lambda\, c^2 \omega_1/(2\Omega^2) + \lambda\, (c^2/2)\, \Omega^{-1} = 0 + O(\lambda^2)$, several other terms cancel out. As a result, the Hamiltonian can then be expressed as

$$H(m_1,\lambda) = \int dp\, \Omega(p)\, \hat{A}^\dagger(p)\, \hat{A}(p) + \lambda\, (c^2/2) \int dp\, \Omega(p)^{-1}\, \delta(0) \tag{4.5}$$

This new Hamiltonian now has the required eigenvalues up to $O(\lambda^1)$, and for the single-particle states the energy is $\omega(p;M) + E_{vac}$, where the (infinite) vaccum's energy is $E_{vac} = \lambda\, (c^2/2) \int dp\, \Omega(p)^{-1}\, \delta(0)$. The second term in Eq. (4.5) can be interpreted as a new interaction, which for the quadratic interaction happens to simply be an infinite constant, but for more complicated interactions it can be a function of the creation and annihilation operators.

**4.2 Alternative mass S-matrix based approach to renormalization**

In contrast to Eq. (4.1), in this approach we assume from the very beginning that the free Hamiltonian is expressed in terms of the correct physical mass M and the "correct" operators $\hat{A}(p)$:

$$H = \int dp\, \omega(p;M)\, \hat{A}^\dagger(p)\hat{A}(p) + \lambda \int dz\, (\hat{\varphi}(z))^2 \tag{4.6}$$

The interaction part of this Hamiltonian, however, is not correct as it violates our required zero-scattering conditions for the vacuum and the single particle states indicated in Eqs. (3.5). In order to "repair" this Hamiltonian we introduce some undetermined counterterm interactions so that the corresponding scattering matrix elements vanish for the vacuum and single particle states. The space of those relevant energy densities that would lead to a quadratic Poincare-invariant interaction can be spanned by the square of the field operator $(\hat{\varphi}(z))^2$ and its covariant derivative $(\partial_{ct}\hat{\varphi}(z))^2 - (\partial_z\hat{\varphi}(z))^2$, so we can try to repair our Hamiltonian with the ansatz

$$H_{phys} \equiv \int dp\, \omega(p;M)\, \hat{A}^\dagger(p)\hat{A}(p) + \lambda \int dz\, (\hat{\varphi}(z))^2 + \int dz\{B(\hat{\varphi}(z))^2 + C(\partial_{ct}\hat{\varphi}(z))^2 - C(\partial_z\hat{\varphi}(z))^2\} \tag{4.7}$$

where we hope that we can determine the two unknown coefficients B and C such that the corresponding first-order scattering matrix elements simultaneously fulfill $\langle 0|S_1|0\rangle = 0$ and $\langle q|S_1|q\rangle = 0$, where $S_1$ is the $O(\lambda)$ contribution to the S-matrix in Eq. (3.4), $S_1 = -i\int_{-\infty}^{\infty} dt\, V(t)$.

Using $\Xi = \int dp/\omega(p)$ we obtain

$$\lambda \Xi + B \Xi + C M^2c^2 \Xi = 0 \qquad (4.8a)$$

$$\lambda [\Xi + 1/(\delta(0) \omega)] + B [\Xi + 1/(\delta(0) \omega)] + C M^2c^2 [\Xi + 1/(\delta(0)\omega)] = 0 \qquad (4.8b)$$

The solutions are $B = -\lambda - M^2c^2 C$ and C is arbitrary. Note that both coefficients are at most linear in $\lambda$. If we insert these coeffcients into the physical Hamiltonian, we obtain

$$H_{phys} = \int dp\, \omega(p;M)\, \hat{A}^\dagger(p)\hat{A}(p) + C (-)\int dp\, \omega(p;M)\, [\hat{A}^\dagger(p)\hat{A}^\dagger(-p)+\hat{A}(p)\hat{A}(-p)] + O(\lambda^2) \qquad (4.9)$$

In other words, the incorrect interaction $\lambda \int dz (\hat{\varphi}(z))^2$ needs to be effectively replaced by the physical interaction $\int dp\, \omega(p)\, [\hat{A}^\dagger(p)\hat{A}^\dagger(-p)+\hat{A}(p)\hat{A}(-p)]$. The free parameter C characterizing the strength of this interaction could then be determined by additional conditions on the system. In this case, it turns out that at second order in perturbation theory C must be set equal to 0, or else it leads to interactions which are not cluster separable. We note that while the repaired Hamiltonian Eq. (4.5) (according to the conventional method) contains an infinite constant in addition to the free Hamiltonian term, the result of this section Eq. (4.9) contains an interaction of the form $\hat{A}^\dagger(p)\hat{A}^\dagger(-p)+\hat{A}(p)\hat{A}(-p)$.

As a side note, we remark that while the quadratic interaction predicts a non-trivial self-scattering of zero and one-particle states, due to its simple structure it has no connected scattering processes between two-particle states. In order to illustrate the alternative renormalization technique for a system with non-trivial two-particle scattering, we study in the next section the more complicated quartic interaction.

## 5. Second example: The quartic interactions

As a simplified model system we choose here the $\hat{\varphi}^4$-interaction [19-22], given by the Hamiltonian

$$H(m,\lambda) = \int dp\, \omega(p;m)\, \hat{a}^\dagger(p;m)\, \hat{a}(p;m) + \lambda \int dz\, (\hat{\varphi}(z))^2 \qquad (5.1)$$

While it can describe many experimentally observable features of critical phenomena, we use it here as a simple but concrete model system to study renormalization [23]. It is our hope that our findings are general enough to also be applicable to the more complicated QED Hamiltonian. In order to have easier access to numerical data we again restrict the spatial dimensions to one.

If we return to the Schrödinger picture representation for the auxilliary field $\hat{\varphi}(z)$, the interaction takes the form:

$$\int dz\, \hat{\varphi}(z)^4 = N[\int dz\, \hat{\varphi}(z)^4] +$$

$$+ (2\pi)\, [(4\pi)^{-1/2} c]^4 \int dp/\omega(p) \int dp_1 \int dp_2\, [\omega(p_1)\omega(p_2)]^{-1/2}$$

$$\{6\, \hat{a}(p_1)\, \hat{a}(p_2)\, \delta(p_1+p_2) + 12\, \hat{a}^\dagger(p_1)\, \hat{a}(p_2)\, \delta(p_2-p_1) + 6\, \hat{a}^\dagger(p_1)\, \hat{a}^\dagger(p_2)\, \delta(p_1+p_2)\} +$$

$$+ (2\pi)\, [(4\pi)^{-1/2} c]^4\, 3\, [\int dp/\omega(p)]^2\, \delta(0) \qquad (5.2)$$

where for notational convenience the five quartic terms associated with the normal-ordered products are abbreviated with

$$N[\int dz\, \hat{\varphi}(z)^4] \equiv (2\pi)\, [(4\pi)^{-1/2} c]^4 \int dp_1 \int dp_2 \int dp_3 \int dp_4\, [\omega(p_1)\omega(p_2)\omega(p_3)\omega(p_4)]^{-1/2}$$

$$\{\hat{a}(p_1)\, \hat{a}(p_2)\, \hat{a}(p_3)\, \hat{a}(p_4)\, \delta(p_1+p_2+p_3+p_4) + 4\, \hat{a}^\dagger(p_1)\, \hat{a}(p_2)\, \hat{a}(p_3)\, \hat{a}(p_4)\, \delta(-p_1+p_2+p_3+p_4) +$$

$$6\, \hat{a}^\dagger(p_1)\, \hat{a}^\dagger(p_2)\, \hat{a}(p_3)\, \hat{a}(p_4)\, \delta(-p_1-p_2+p_3+p_4) + 4\, \hat{a}^\dagger(p_1)\, \hat{a}^\dagger(p_2)\, \hat{a}^\dagger(p_3)\, \hat{a}(p_4)\, \delta(-p_1-p_2-p_3+p_4) +$$

$$\hat{a}^\dagger(p_1)\, \hat{a}^\dagger(p_2)\, \hat{a}^\dagger(p_3)\, \hat{a}^\dagger(p_4)\, \delta(-p_1-p_2-p_3-p_4)\} \qquad (5.3)$$

Furthermore, if we replace the infinite upper and lower integration limits with the cut-off momentum $\Lambda$, the integral $\int dp/\omega(p)$ in Eq. (5.2) can be evaluated as $(2/c)\, \mathrm{Sinh}^{-1}(\Lambda/mc) \equiv \Xi$. As it will be relevant for the discussion below, we point out that the interaction can also be expressed as

$$\int dz\, \hat{\varphi}(z)^4 = N[\int dz\, \hat{\varphi}(z)^4] + 6[(4\pi)^{-1/2} c]^2\, \Xi \int dz\, \hat{\varphi}(z)^2 - 3(2\pi)\, [(4\pi)^{-1/2} c]^4\, \Xi^2\, \delta(0)$$

$$= N[\int dz\, \hat{\varphi}(z)^4] + 6[(4\pi)^{-1/2} c]^2\, \Xi\, N[\int dz\, \hat{\varphi}(z)^2] + 3(2\pi)\, [(4\pi)^{-1/2} c]^4\, \Xi^2\, \delta(0) \qquad (5.4)$$

Here we have used an interesting equality that shows that by adding a term $\int dz \hat{\varphi}(z)^2$ to $\int dz\, \hat{\varphi}(z)^4$ the resulting expression contains only normal ordered terms:

$$\int dz \hat{\varphi}(z)^4 - 3[(4\pi)^{-1/2}c]^2 \, \Xi \int dz \hat{\varphi}(z)^2 = N[\int dz \hat{\varphi}(z)^4] + 3[(4\pi)^{-1/2}c]^2 \, \Xi \, N[\int dz \hat{\varphi}(z)^2] \qquad (5.5)$$

## 5.1 The role of single- and two-particle masses

Let us assume that as a result of a measurement of our system, we know that the mass of a single particle has the specific value denoted by $M_1$. Furthermore, let us assume that due to another measurement (via some cross-section, e.g.) we also know that the interaction strength between two particles is such that the lowest energy of a state with two particles takes the specifc value of $M_2 c^2$. For the special case of no interaction ($\lambda=0$), we would trivially find that the two-particle energy eigenstate is just the direct product of two single-particle states, $\hat{a}^\dagger(p_1)\hat{a}^\dagger(p_2)|vac\rangle$ with the eigenvalue $\omega(p_1)+\omega(p_2)$. In this special (non-interacting) case, the lowest energy eigenvalue $M_2 c^2$ would be equal to $2M_1 c^2$, associated with $p_1=p_2=0$. In other words, the effective two-particle mass is simply the sum of two single masses. However, if the two particles interact with each other in an attractive way then the lowest possible energy can be less than $2M_1 c^2$. The true size of $M_2$ is therefore a measure for the strength of the interaction. This "two-particle mass" $M_2$ is related to the charge (or coupling strength) whose size determines the magnitude of the interaction. If the original Hamiltonian $H_0 + \lambda \int dz \, \hat{\varphi}(z)^4$ was a correct description of our system, it would have to reproduce the two desired values $M_1 c^2$ and $M_2 c^2$ as its lowest eigenvalues associated with the single- and two-particle sector. Furthermore, as its vacuum state is supposed to be free of any particles, we would also expect its lowest energy eigenvalue to be zero. However, H has none of these desired spectral properties in its present form.

A closer inspection of the Hamiltonian $H=H_0+\lambda \int dz \, \hat{\varphi}(z)^4$ shows that it can be re-expressed as the sum of normal ordered products plus an infinite constant which we denote by $\Theta \equiv \lambda\, 3(2\pi) [(4\pi)^{-1/2}c]^4 \, \Xi^2 \, \delta(0)$, which depends on the bare mass m through $\Xi=\int dp/\omega(p)$. For a system in a finite box of length L, the $\delta(0)$ should be replaced by $\delta(0)=L/(2\pi)$. We will return to this in our numerical discussion of Section 5.3 below. In order to renormalize our Hamiltonian by finding the parameters m and $\lambda$, we first subtract the constant $\Theta$ from it. This subtraction changes the spectrum in a trivial way and leaves all eigenvectors the same, and gives the new interaction $V \equiv \lambda \int dz\ \hat{\varphi}(z)^4 - \Theta = \lambda N[\int dz\ \hat{\varphi}(z)^4] + 6\lambda[(4\pi)^{-1/2}c]^2 \Xi\ N[\int dz\ \hat{\varphi}(z)^2]$ the necessary vanishing vacuum expectation value, as discussed in Appendix A. This interaction V now consists entirely of normal ordered products.

In order to repair (renormalize) the new Hamiltonian H, we can compute the spectrum of it and adjust the two free parameters m and $\lambda$, until the spectrum matches the two renormalization conditions, here given by $\min(\{E(1)\}) = M_1 c^2$ and $\min(\{E(2)\}) = M_2 c^2$, where the set $\{E(n)\}$ are the eigenvalues of H in sectors with n physical particles.

### 5.1.1 Single-particle mass using Rayleigh–Schrödinger perturbation theory

In usual (non-degenerate) Rayleigh-Schrödinger perturbation theory [24] the energies $\Omega$ of the Hamiltonian $H=H_0+V$ can be computed as the sum $\Omega = \Sigma_n\ \omega^{(n)}$, where the general form of the two lowest terms is given by $\omega^{(1)} = \langle\omega^{(0)}|V|\omega^{(0)}\rangle/\langle\omega^{(0)}|\omega^{(0)}\rangle$ and $\omega^{(2)} = \langle\omega^{(0)}|\omega^{(0)}\rangle^{-1} \Sigma_\beta |\langle\beta^{(0)}|V|\omega^{(0)}\rangle|^2 (\omega^{(0)} - \beta^{(0)})^{-1}$. Here the summation (integration) $\Sigma_\beta$ extends over all eigenstates $|\beta^{(0)}\rangle$ of $H_0$ with energy $\beta^{(0)}$, except the state $|\omega^{(0)}\rangle$.

If we restrict our scheme to only the lowest order in $\lambda$, the two renormalization conditions simplify to (1) $\omega(p)|_{p=0} + \langle p|V|p\rangle/\delta(0)|_{p=0} = M_1 c^2$ and (2) $2\omega(p)|_{p=0} + \langle p,p|H|p,p\rangle/\delta(0)^2|_{p=0} = M_2 c^2$. In other words, we simply have to compute the three diagonal elements of the potential and invert them to find m and $\lambda$ as a function of the given values $M_1$ and $M_2$.

To incorporate the first requirement we have to compute the energy of the diagonal element for the single particle state $\langle p|V|p\rangle$, which amounts to

$$\langle p|\ V\ |p\rangle = \lambda\ \langle p|\int dz\ \hat{\varphi}(z)^4 - \Theta\ |p\rangle$$

$$= \lambda\ \langle p|(2\pi)\ [(4\pi)^{-1/2}c]^4\ \Xi\ \int dp_1 \int dp_2\ [\omega(p_1)\omega(p_2)]^{-1/2}\ 12\ \hat{a}^\dagger(p_1)\ \hat{a}(p_2)\ \delta(p_2 - p_1)\ |p\rangle$$

$$= \lambda\,(2\pi)\,[(4\pi)^{-1/2}c]^4\,\Xi\,\omega(p)^{-1}\,12\,\delta(0) \tag{5.6}$$

We obtain $E_p(\lambda) = \omega(p) + \lambda\,(2\pi)\,[(4\pi)^{-1/2}c]^4\,\Xi\,\omega(p)^{-1}\,12 + O(\lambda^2)$, which we need to evaluate for p=0 to obtain the mass, according to our requirement $E_{p=0}(\lambda)= M_1 c^2$. Using $\Xi=(2/c)\sinh^{-1}(\Lambda/mc)$ we find

$$M_1 = m + \lambda\,3(\pi m c)^{-1}\sinh^{-1}(\Lambda/mc) + O(\lambda^2) \tag{5.7}$$

which is a transcendental equation for m as a function of the required mass $M_1$. Consistent with our perturbative approach, we can solve this equation up to order $O(\lambda)$ and find for our renormalized mass m

$$m = M_1 - \lambda\,3(\pi M_1 c)^{-1}\sinh^{-1}(\Lambda/M_1 c) + O(\lambda^2) \tag{5.8}$$

In the limit of $\Lambda \to \infty$, we can replace $\sinh^{-1}$ by its asymptotic behavior $\sinh^{-1}(\Lambda/M_1 c) \to \ln[2\Lambda/M_1 c] + (1/4)(\Lambda/M_1 c)^{-2}$.

The first-order renormalized mass can then be used to compute the second-order correction to the mass. The second-order contribution to the mass shift can be calculated from the second-order perturbative correction to the energy

$$E^{(2)} = \delta(0)^{-1} \int d\beta\,|\langle\beta|V|p\rangle|^2 / (\omega_p - \omega_\beta) \tag{5.9}$$

where $\int d\beta$ is the integration over all possible intermediate states and the factor of $\delta(0)^{-1}$ arises from the normalization of the single particle state, $\langle p|p\rangle = \delta(0)$. The two terms in the interaction V which modify the energy of a single particle state $|p\rangle$ at second order in $\lambda$ are proportional to $\hat{a}^\dagger \hat{a}^\dagger \hat{a} \hat{a}$, corresponding to $|\beta\rangle$ being a 3-particle state, and $\hat{a}^\dagger \hat{a}^\dagger \hat{a}^\dagger \hat{a}^\dagger$, where $|\beta\rangle$ will be a 5-particle state. We will define the two corresponding second order energy corrections as $E^{(2)}_3(p)$ and $E^{(2)}_5(p)$. The details are worked out in Appendix B. The final contribution to the single-particle energy $E^{(2)}(p)$ is

$$E^{(2)}(p) = E^{(2)}_3(p) + E^{(2)}_5(p) - E^{(2)}_{vac} \qquad (5.10)$$

$$= 3/(2\pi^2)\, c^8 \lambda^2 \iint dp_1 dp_2\, [\omega(p_1)\omega(p_2)\omega(p-p_1-p_2)\omega(p)]^{-1} / (\omega_p - \omega(p_1) - \omega(p_2) - \omega(p-p_1-p_2))$$

$$+ 3/(2\pi^2)\, c^8 \lambda^2 \iint dp_1 dp_2\, [\omega(p_1)\omega(p_2)\omega(-p-p_1-p_2)\omega(p)]^{-1} /(-\omega(p_1) - \omega(p_2) - \omega(-p-p_1-p_2) - \omega(p))$$

where $E^{(2)}_{vac}$ is the energy of the vacuum state itself, which must be subtracted from the single-particle state's energy in order to isolate that part of the state's energy which corresponds to the shift in the particle's mass due to the interaction. The correction to the mass can be evaluated by evaluating $E^{(2)}(p=0)$ numerically; for instance, for a bare mass of m=1, this mass correction is $-\lambda^2 \times 0.75$. This result agrees with the result which will be calculated from Feynman graphs in Section 5.1.2.

In order to make contact with computational methods, to be presented in Section 5.1.4, we can discretize the momentum by confining the system to a finite box with periodic boundary conditions [25]. For a box of length L the momentum modes of the system are discrete with spacing $\Delta=2\pi/L$. We can therefore replace all momentum integrals $\int dp$ by $\Delta \Sigma_p$. The creation and annihilation operators must be normalized by $\sqrt{(\Delta)}$, $\hat{a}_p \equiv \sqrt{(\Delta)}\, \hat{a}(p)$ and $\hat{a}_p^\dagger \equiv \sqrt{(\Delta)}\, \hat{a}^\dagger(p)$, where now the subscript p is an integer and the physical momentum corresponding to integer momentum p is given by $p\Delta$. This normalization results in the normalized creation and annihilation operators satisfying the commutation relations $[\hat{a}_p, \hat{a}_{p'}^\dagger] = \delta_{pp'}$, where $\delta_{pp'}$ is the Kronecker delta.

With these substitutions, the second order perturbative correction to the energy of a single particle state from the term $\hat{a}^\dagger \hat{a}^\dagger \hat{a}^\dagger \hat{a}$ for this discretized system is

$$E^{(2)Dis}_3(k) = \lambda^2/96\, c^{10}/L^2\, \Sigma_{k1}\Sigma_{k2}\, (\omega_{k1}\omega_{k2}\omega_{k-k1-k2}\omega_k)^{-1}\, (\omega_k - \omega_{k1} - \omega_{k2} - \omega_{k-k1-k2})^{-1} \qquad (5.11)$$

while the corrrection term from 5-particle intermediate states, resulting from the term $\hat{a}^\dagger \hat{a}^\dagger \hat{a}^\dagger \hat{a}^\dagger$, is

$$E^{(2)\,Dis}_{5}(k) = \lambda^2/96\ c^{10}/L^2\ \Sigma_{k1}\Sigma_{k2}\ (\omega_{k1}\omega_{k2}\omega_{k+k1+k2}\omega_k)^{-1}\ (-\omega_k-\omega_{k1}-\omega_{k2}-\omega_{k+k1+k2})^{-1} \quad (5.12)$$
$$+ \lambda^2/384\ c^{10}/L^2\ \Sigma_{k1}\Sigma_{k2}\Sigma_{k3}\ (\omega_{k1}\omega_{k2}\omega_{k3}\omega_{-k1-k2-k3})^{-1}\ (-\omega_{k1}-\omega_{k2}-\omega_{k3}-\omega_{k+k1+k2})^{-1}$$

The second term in this equation is equal to the second order correction term of the vacuum state $|0\rangle$ for a discrete system,

$$E^{(2)\,Dis}_{vac} = \lambda^2/384\ c^{10}/L^2\ \Sigma_{k1}\Sigma_{k2}\Sigma_{k3}\ (\omega_{k1}\omega_{k2}\omega_{k3}\omega_{-k1-k2-k3})^{-1}\ (-\omega_{k1}-\omega_{k2}-\omega_{k3}-\omega_{k+k1+k2})^{-1} \quad (5.13)$$

In the limit of $L\to\infty$, these discretized results approach the continuum limit of Eq. (5.10), as well as agreeing with the result of Feynman diagram calculations in Section 5.1.2 below.

### 5.1.2 Single-particle mass using the usual propagator based Feynman technique

In this particular approach, the physical single particle mass is determined from the two-point correlation function of the fields by assuming that the functional dependence of the momentum is proportional to $(p^2-M^2)^{-1}$.

To first order in $\lambda$, the only diagram which contributes to the two-point function is Figure 1. This diagram (and all other diagrams containing it as a subdiagram) may be removed from the theory by normal-ordering the Hamiltonian, since no two fields inside of a normal-ordered interaction term will Wick contract on each other and produce loops as in Figure 1.

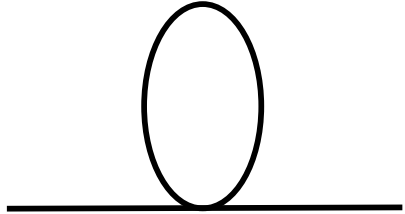

Figure 1. $O(\lambda)$ correction to the two-point correlation function.

The $O(\lambda^2)$ contribution is the two-loop diagram Figure 2, and is given by

$$-i\,\Sigma(k^2) = (-24i\lambda)^2/6 \int d^2q/(2\pi)^2 \int d^2p/(2\pi)^2\ i/(p^2-m^2)\ i/(q^2-m^2)\ i/((k-p-q)^2-m^2) \quad (5.14)$$

```
        p
      k-p-q
   k       k
```

The two momentum integrals can be completed by introducing Feynman parameters and then Wick rotating the energy integrations. The result is

$$\Sigma(k^2) = -6\lambda^2/\pi^2 \int_0^1 dx \int_0^1 dy \int_0^1 dz\, \delta(x+y+z-1) / ((x^2z^2 - xz(xy+xz+yz))/(x+z)\, k^2 + (xy+xz+yz)\, m^2)$$

(5.15)

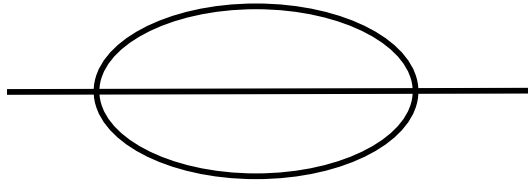

Figure 2. The $O(\lambda^2)$ contribution to the two-point function.

The two-point correlation function, accurate to $O(\lambda^2)$, can be written as $i/(p^2 - m^2 - \Sigma(k^2))$. The true mass of the particle is given by the poles in the two-point function, and is therefore defined by the transcendental equation $M^2 = m^2 - \Sigma(M^2)$. To $O(\lambda^2)$, this can be written as $M^2 = m^2 - \Sigma(m^2)$. The integrals in Eq. (5.15) can be evaluated numerically at $k^2 = m^2$, and for a bare mass of m=1 the result is $\Sigma((m=1)^2) \approx -\lambda^2 \times 1.51$. The renormalized mass is therefore $M = \sqrt{(m^2 - \Sigma(m^2))} = m\sqrt{(1 - \Sigma(m^2)/m^2)} \approx m - 1/2\, \Sigma(m^2)/m \approx 1 - \lambda^2 \times 0.75$, in agreement with the mass correction calculated in Section 5.1.2.

### 5.1.3 Two-particle mass using perturbation theory

To incorporate the no-self-scattering requirements we have to compute the energy of the diagonal element for the two-particle state $\langle p,p|V|p,p\rangle$, which contains two terms associated with the interactions proporational to $\hat{a}^\dagger \hat{a}$ and $\hat{a}^\dagger \hat{a}^\dagger \hat{a}\, \hat{a}$

$$\langle p,p|\, V\, |p,p\rangle = \lambda\, \langle p,p| \int dz\, \hat{\varphi}(z)^4 - \Theta\, |p,p\rangle$$

19      10/10/12

$$= \lambda \langle p,p|(2\pi) [(4\pi)^{-1/2}c]^4 \ \Xi \int dp_1 \int dp_2 \ [\omega(p_1)\omega(p_2)]^{-1/2} \ 12 \ \hat{a}^\dagger(p_1) \ \hat{a}(p_2) \ \delta(p_2-p_1) \ |p,p\rangle$$

$$+ \lambda \langle p,p| (2\pi) [(4\pi)^{-1/2}c]^4 \int dp_1 \int dp_2 \int dp_3 \int dp_4 \ [\omega(p_1)\omega(p_2)\omega(p_3)\omega(p_4)]^{-1/2}$$

$$6 \ \hat{a}^\dagger(p_1) \ \hat{a}^\dagger(p_2) \ \hat{a}(p_3) \ \hat{a}(p_4) \ \delta(p_3+p_4-p_1-p_2) \ |p,p\rangle \tag{5.16}$$

Using $\langle p,p| \hat{a}^\dagger(p_1) \hat{a}(p_2) |p,p\rangle = \delta(p_1-p) \delta(p_2-p) \sqrt{2} \ \delta(0)$ and $\langle p,p|\hat{a}^\dagger(p_1)\hat{a}^\dagger(p_2)\hat{a}(p_3)\hat{a}(p_4)|p,p\rangle = \delta(p_1-p) \delta(p_2-p)\delta(p_3-p) \delta(p_4-p) \sqrt{2} \ \delta(0)$, the integrals can be performed, leading to

$\langle p,p| V|p,p\rangle = \lambda \ (2\pi) [(4\pi)^{-1/2}c]^4 \ \Xi \ [\omega(p)]^{-1} \ 12 \sqrt{2} \ \delta(0)^2 + \lambda \ (2\pi) [(4\pi)^{-1/2}c]^4 \ [\omega(p)]^{-2} \ 6 \sqrt{2} \ \delta(0)^2$

such that the two-particle mass becomes

$$M_2 c^2 = 2 \ \omega(0) + \lambda \ 6 \sqrt{2} /2 \ (4\pi)^{-1} c^4 \ \{2 \ \Xi \ m^{-1} c^{-2} + m^{-2} c^{-4}\} \tag{5.16}$$

This quantity, being infinite, must be renormalized to the correct value. This is accomplished by including in the theory a counterterm of the form $C \int dz \hat{\varphi}(z)^4 - C \ 3(2\pi) [(4\pi)^{-1/2}c]^4 \ \Xi^2 \ \delta(0)$, where once again we have subtracted an infinite constant from the counterterm so that it is equal to normal-ordered products of operators, which have zero vacuum expectation value. With the addition of this counterterm, the two-particle mass becomes

$$M_2 c^2 = 2 \ \omega(0) + (\lambda+C) \ 6 \sqrt{2} /2 \ (4\pi)^{-1} c^4 \ \{2 \ \Xi \ m^{-1} c^{-2} + m^{-2} c^{-4}\} \tag{5.17}$$

The value of C must be chosen to be an infinite constant, such that the two-particle mass becomes finite and equal to the known, measured value.

**5.1.4 Complete spectrum using direct compuational methods**

In this approach we try to find an efficiently truncated set of basis states that permits us to determine the energy eigenvalues directly by diagonalizing the corresponding Hamiltonian matrix numerically [25,26]. If we asume that the system is inside a box of length L with periodic boundary conditions, we can discretize the spatial axis into (2N+1) grid points. As a result, the discretized momenta take the form $p_j=j(2\pi)/L$, where $j=0,\pm 1, \pm 2, \pm N$ and the largest momentum is $\Lambda = N(2\pi)/L$. We have used the discretized versions of the energy eigenstates of $H_0$ as basis states.

After discretization, the system is still infinite dimensional as there can be an arbitrary number of bosons in each state. Thus, the system must be further restricted by setting an upper limit $N_b$ to the total number of bosons. Due to momentum conservation each momentum block can be diagonalized separately, and the largest momentum block (generally the zero momentum subspace) can, due to memory limitations, be at most a $10^5 \times 10^5$ Hamiltonian matrix. Diagonalization can be further simplified, due to the $\hat{\varphi} \to -\hat{\varphi}$ symmetry, such that states with an even number of bosons do not couple with states that have odd boson numbers, which allows each of these subspaces to be considered separately. With recently developed computational techniques based on dynamically allocated pointer variables and suitable memory swapping, one can now simulate the fully coupled quantum field theoretical fermion-boson interaction in less than 50 CPU-hours of computing time.

If the maximum allowed momentum mode is $N=10$ and the number of allowed bosons is $N_b=5$, the zero momentum subspace with an odd number of bosons is a $1456 \times 1456$ matrix which must be diagonalized. For coupling constant $\lambda=0.01$, which is well inside the perturbative regime, and for box length $L=0.02$, the lowest lying eigenvalue is $18778.843$ in atomic units where $c=137.036$ a.u., which when compared to the bare energy of a boson of zero momentum $c^2=18778.865$, yields an energy correction $\Delta E = -0.022$. This result is to be compared with the results for discretized perturbation theory, Eqs. (5.11)–(5.13). Eq. (5.11) gives $E^{(2)}_{3}{}^{Dis}(k) = -0.012$ for these parameters, while from Eq. (5.12) we have $E^{(2)}_{5}{}^{Dis}(k) = -0.010$ for a total energy correction of $-0.022$, identical to the numerical results. We should note that Eq. (5.12) suggests that the perturbative correction to the energy of the single particle state contains a term identical to the energy correction of the vacuum state. This change to the vacuum energy should be subtracted from the state's energy, and whatever energy correction remains is the true change to the particle's rest energy. For these parameters, the vacuum energy correction is $E^{(2)}_{vac}{}^{Dis} = -0.003$, so the actual change to the particle's rest energy is $\Delta mc^2 = -0.019$.

This numerical method of solving for the spectrum of a quantum field theory can in principle be used to renormalize such a theory non-perturbatively. The restriction of the number of momentum modes to some maximum N, which was necessary in order to make the number of modes finite so that the theory becomes numerically tractable, also automatically regulates the integrals by playing the role of a momentum cutoff. The bare parameters can then be adjusted until the observables, such as the rest energy $E=mc^2$, match their physical values. These bare parameters would depend on both the box length and the maximum momentum mode N, since the momentum cutoff is given by the momentum of the highest mode, which is $\Lambda = N\, 2\pi/L$. This procedure could be carried out iteratively, and since the numerical method is non-perturbative the value of the bare parameters derived from this procedure would be the exact, non-perturbative parameters that are necessary in order to yield the correct physical parameters.

**5.2 Alternative S-matrix based approach to renormalization**

In this alternative renormalization approach, we have to construct the counterterms $V_{counter}(z)$ for the Hamiltonian $H = \int dp\, \Omega(p)\, \hat{A}^\dagger(p)\, \hat{A}(p) + \lambda \int dz\, \hat{\varphi}(z)^4$ where $\Omega(p) \equiv [M^2c^4 + c^2p^2]^{1/2}$, such that the no-self-scattering conditions of Eqs. (3.5) are fulfilled. In full generality we start here with the ansatz

$$V_{counter}(z) = B\int dz\, \hat{\varphi}(z)^2 + C \int dz\, \hat{\varphi}(z)^4 + D \int dz\, D(\hat{\varphi}) + E \int dz\, D(\hat{\varphi})^2 \qquad (5.19)$$

where $D(\hat{\varphi}) = (\partial_{ct}\hat{\varphi}(z))^2 - (\partial_z\hat{\varphi}(z))^2$. The four unknown coefficients will be determined below.

The perturbative expansion (3.4) of the S matrix is given by $S = \sum_{n=0}^{\infty} S_n$, where $S_n \equiv (-i)^n /n!\int ...\int dt_1 dt_2 .. dt_n\, T\{V(t_1)V(t_2)...V(t_n)\}$. Note that we have introduced the Dyson expansion form, which is equivalent to the expansion Eq. (4.10), but because all integrals extend from $-\infty$ to $\infty$, it requires the introduction of the time-ordering operator T.

In particular, we compute the matrix elements for the vacuum first, and they can be easily computed if we insert the mode decomposition of the field Eq. (2.2) into the corresponding equations. After a lengthy calculation we obtain the rather simple results:

$$\int dz\, \langle 0|(\hat{\varphi}(z,t))^2|0\rangle = c^2/2\, \delta(0)\, \Xi \qquad (5.20a)$$

$$\int dz\, \langle 0|(\hat{\varphi}(z,t))^4|0\rangle = 3(8\pi)^{-1} c^4\, \Xi^2\, \delta(0) \qquad (5.20b)$$

$$\int dz\, \langle 0|(\partial_{ct}\hat{\varphi}(z))^2 - (\partial_z\hat{\varphi}(z))^2|0\rangle = (m^2c^4/2)\, \Xi\, \delta(0) \qquad (5.20c)$$

$$\int dz\, \langle 0|\{(\partial_{ct}\hat{\varphi}(z))^2 - (\partial_z\hat{\varphi}(z))^2\}^2|0\rangle = (8\pi)^{-1} [m^2c^4\Xi]^2\, \delta(0)$$

$$+ (4\pi)^{-1} \{[\int dp_1 \, \omega(p_1)]^2 + [\int dp \, c^2 p^2/\omega(p)]^2\} \, \delta(0) \qquad (5.20d)$$

In Appendix C we show that the corresponding values for the single-particle particle sector can be obtained directly from these expressions if we use the general relationships:

$$\langle q|(\hat{\varphi}(z,t))^N|q\rangle = \langle 0|(\hat{\varphi}(z,t))^N|0\rangle \, \delta(0) + N(N-1) \, c^2 (4\pi)^{-1} \, \omega(q)^{-1} \, \langle 0|\hat{\varphi}(z,t)^{N-2}|0\rangle \qquad (5.21)$$

$$\langle q|(D\hat{\varphi})^N|q\rangle = \langle 0|(D\hat{\varphi})^N|0\rangle \, \delta(0) + 2N(4\pi)^{-1} c^4 m^2/\omega(q) \langle 0|(D\hat{\varphi})^{N-1}|0\rangle +$$
$$+ 4N(N-1)(4\pi)^{-1} \omega(q)^{1/2} \langle 0|(\partial_{ct}\hat{\varphi}(z) + cq/\omega(q) \, \partial_z\hat{\varphi}(z))^2 (D\hat{\varphi})^{N-2}|0\rangle \qquad (5.22)$$

We provide the proofs for these useful relationships in Appendix C. Applied to our operators, we obtain

$$\int dz \, \langle q|(\hat{\varphi}(z,t))^2|q\rangle = c^2/2 \, \delta(0) \, \Xi \, \delta(0) + c^2 \, \omega(q)^{-1} \, \delta(0) \qquad (5.23a)$$

$$\int dz \, \langle q|(\hat{\varphi}(z,t))^4|q\rangle = 3(8\pi)^{-1} c^4 \Xi^2 \, \delta(0) \, \delta(0) + 3(2\pi)^{-1} c^4 \, \omega(q)^{-1} \, \delta(0) \, \Xi \qquad (5.23b)$$

$$\int dz \, \langle q|(\partial_{ct}\hat{\varphi}(z))^2 - (\partial_z\hat{\varphi}(z))^2|q\rangle = (m^2 c^4/2) \, \Xi \, \delta(0) + c^4 m^2/\omega(q) \, \delta(0) \qquad (5.23c)$$

$$\int dz \, \langle q|\{(\partial_{ct}\hat{\varphi}(z))^2 - (\partial_z\hat{\varphi}(z))^2\}^2|q\rangle = (8\pi)^{-1} [m^2 c^4 \Xi]^2 \, \delta(0) \, \delta(0) +$$
$$+ (4\pi)^{-1} \{[\int dp_1 \, \omega(p_1)]^2 + [\int dp \, c^2 p^2/\omega(p)]^2\} \, \delta(0) \, \delta(0) + (\pi)^{-1} c^4 m^2/\omega(q) \, (m^2 c^4/2) \, \Xi \, \delta(0) +$$
$$+ (\pi)^{-1} [\omega(q) \int dp \, \omega(p) + c^2 q^2 \omega(q)^{-1} \int dp \, c^2 p^2 \, \omega(p)^{-1}] \, \delta(0) \qquad (5.23d)$$

Using these relations, we can calculate the perturbative corrections to the S-matrix and then require that the vacuum and single particle states satisfy the no-self-scattering conditions order by order in perturbation theory. Since the zeroth-order (free) perturbative result already satisfies $\langle 0|S_0|0\rangle = 1$ and $\langle q|S_0|q\rangle = \delta(0)$, the no-self-scattering condition reduces to the requirement that the higher order perturbative corrections must have vanishing expectation values in the vacuum and single particle states. The system is underdetermined, as for instance for the interaction $V(z) + V_{counter}(z)$ with $V(z) = \lambda \int dz \, \hat{\varphi}(z)^4$, there are 4 unknown coefficients in $V_{counter}(z)$ and only the two equations for the vanishing of the vacuum and single particle expectation values. To first order in perturbation theory, the equation associated with the vacuum is

$$0 = \langle 0|S_1|0\rangle$$

$$= -i \int dt \, [B \, c^2/2 \, \Xi \, \delta(0) + (C+\lambda) \, 3(8\pi)^{-1} \, c^4 \Xi^2 \, \delta(0) + D \, (m^2 c^4/2) \, \Xi \, \delta(0)$$

$$+ E \, [(8\pi)^{-1} [m^2 c^4 \Xi]^2 \delta(0) + (4\pi)^{-1} \{[\int dp_1 \, \omega(p_1)]^2 + [\int dp \, c^2 p^2/\omega(p)]^2\} \delta(0)] \,]$$

$$= B \, c^2/2 \, \Xi + (C+\lambda) \, 3(8\pi)^{-1} \, c^4 \Xi^2 + D \, (m^2 c^4/2) \, \Xi \tag{5.24}$$

$$+ E \, [(8\pi)^{-1} [m^2 c^4 \Xi]^2 + (4\pi)^{-1} \{[\int dp_1 \, \omega(p_1)]^2 + [\int dp \, c^2 p^2/\omega(p)]^2\}]$$

while the equation coming from the the no-self-scattering condition applied to the single particle states is

$$0 = \langle q|S_1|q\rangle$$

$$= -i \int dt \, [B \, [c^2/2 \, \delta(0) \, \Xi \, \delta(0) + c^2 \, \omega(q)^{-1} \, \delta(0)]$$

$$+ (C+\lambda) \, [3(8\pi)^{-1} \, c^4 \, \Xi^2 \, \delta(0) \, \delta(0) + 3(2\pi)^{-1} \, c^4 \, \omega(q)^{-1} \, \delta(0) \, \Xi \,]$$

$$+ D \, [(m^2 c^4/2) \, \Xi \, \delta(0) \, \delta(0) + c^4 m^2/\omega(q) \, \delta(0)]$$

$$+ E \, [(8\pi)^{-1} \, [m^2 c^4 \Xi]^2 \, \delta(0) \, \delta(0) + (4\pi)^{-1} \, \{[\int dp_1 \, \omega(p_1)]^2 + [\int dp \, c^2 p^2/\omega(p)]^2\} \, \delta(0) \, \delta(0)$$

$$+ (\pi)^{-1} c^4 m^2/\omega(q)(m^2 c^4/2) \Xi \, \delta(0) + (\pi)^{-1} [\omega(q) \int dp \, \omega(p) + c^2 q^2 \omega(q)^{-1} \int dp \, c^2 p^2 \, \omega(p)^{-1}] \delta(0)]$$

$$= B \, [c^2/2 \, \Xi \, \delta(0) + c^2 \, \omega(q)^{-1}] + (C+\lambda) \, [3(8\pi)^{-1} \, c^4 \, \Xi^2 \, \delta(0) + 3(2\pi)^{-1} \, c^4 \, \omega(q)^{-1} \, \Xi \,]$$

$$+ D \, [(m^2 c^4/2) \, \Xi \, \delta(0) + c^4 m^2/\omega(q)]$$

$$+ E \, [(8\pi)^{-1} \, [m^2 c^4 \Xi]^2 \, \delta(0) + (4\pi)^{-1} \, \{[\int dp_1 \, \omega(p_1)]^2 + [\int dp \, c^2 p^2/\omega(p)]^2\} \, \delta(0)$$

$$+ (\pi)^{-1} c^4 m^2/\omega(q)(m^2 c^4/2) \Xi + (\pi)^{-1} [\omega(q) \int dp \, \omega(p) + c^2 q^2 \omega(q)^{-1} \int dp \, c^2 p^2 \, \omega(p)^{-1}] \,]$$

(5.25)

We may solve these equations for B and E in terms of the other unknown coefficients C and D. Doing so yields

$$B = -D \, m^2 c^2 - 3/(4\pi) \, (C+\lambda) \, c^2 \Xi \tag{5.26a}$$

$$+ (6 (C+\lambda) c^2 \Xi) (m^4 c^8 \Xi^2/(8\pi) + 1/(4\pi)\{[\int dp\, \omega(p)]^2 + c^4[\int dp\, p^2/\omega(p)]^2\})$$

$$\times (m^4 c^8 \Xi^2 + 4 q^2 c^4 \Xi \int dp\, p^2/\omega(p) - 2 c^4 [\int dp\, p^2/\omega(p)]^2 + 4 \Xi \omega(q)^2 \int dp\, \omega(p))^{-1}$$

$$E = - 3 (C+\lambda) c^4 \Xi^2) \quad (5.26b)$$

$$\times (m^4 c^8 \Xi^2 + 4 q^2 c^4 \Xi \int dp\, p^2/\omega(p) - 2 c^4 [\int dp\, p^2/\omega(p)]^2 + 4 \Xi \omega(q)^2 \int dp\, \omega(p))^{-1}$$

The remaining coefficients of the counterterms C and D must be determined by other physical requirements. One such requirement is that when these counterterms are applied to higher orders in perturbation theory that the S-matrix remains cluster separable. Also, the physical mass of the particle and its measured coupling constant will set further constraints on these coefficients and thereby determine the counterterms.

## 6. Discussion and open questions

In the standard field theoretical treatment of renormalization, the field theory is renormalized based on the requirement that the mass and coupling match their measured values and the field strength is renormalized by requiring that the field operator creates single particle states, as assumed by the LSZ reduction formula. But the renormalization program can also be carried out by applying a no-self-scattering condition on the S-matrix, so that the vacuum and single particle states are stable and do not scatter with themselves. Both of these techniques were applied to both $\hat{\varphi}(z)^2$ and $\hat{\varphi}(z)^4$ interactions.

In the standard treatment of renormalization, there are two renormalization conditions for each particle species in the theory: mass renormalization and wave function renormalization. In the S-matrix renormalization presented here, these two conditions are seemingly replaced by the sole condition that single particle states should not be self-scattering. This raises interesting questions as to whether this renormalization condition is complete, that is, whether it contains enough conditions to fully determine the parameters of the theory. It is interesting to note that we had to employ the additional condition of cluster separability; whether cluster separability is sufficient to fix all possible parameters at all higher orders in perturbation theory remains an open question.

Further work must also be done to include coupling constant renormalization into this method. The issue was briefly studied here by examining the two-particle mass, which is the energy of the lowest lying two-particle eigenstate divided by $c^2$. Such a quantity may make an adequate assessment of the coupling strength for use in the numerical techniques presented here, since our numerical method involves the diagonalization of the Hamiltonian which directly gives the energies of all states, including two particle states. However, the two-particle mass of a bound state is obviously not a good measurement to use for renormalization within the context of S-matrices, as bound states are not easily handled in S-matrices and are fundamentally non-perturbative.

Quantum field theories can also be renormalized using numerical techniques. The discretization necessary to make a numerical solution to a quantum field theory feasible automatically regulates the integrals, and the bare mass and coupling could be adjusted iteratively until they match the desired physical parameters. Since the numerical method employed here is a non-perturbative solution to the $\hat{\varphi}(z)^4$ theory, such a renormalization scheme promises to provide a non-perturbative technique for the regularization and renormalization of any quantum field theory, incuding QED. Such a non-perturbative solution to QFT has yet to be realized, partly due to computational limitations and partly due to conceptual difficulties, but we hope that the current study may be a starting point for this line of research.

Although the techniques presented here were applied to the simple, concrete examples of $\hat{\varphi}(z)^2$ and $\hat{\varphi}(z)^4$ theories in one spatial dimension, there such methods may be applicable to theories of higher spin in higher dimension, including the Yukawa theory or QED. In QED, the no-self-scattering conditions could be applied to three states: the vacuum, the single photon state, and the single electron state. Other physical conditions such as cluster separability provide further constraints on the parameters used in the counterterms.

There remain many open questions, which must be reserved for future works. One such question is whether the S-matrix renormalization method presented here can be shown to work at all orders of perturbation theory; ideally, one would like to have a proof that the theory is renormalizable order-by-order at all orders of the coupling constant. A further question remains as to how to apply numerical techniques to a gauge theory due to the gauge invariance. In particular, numerical methods naturally lend themselves to a maximum momentum cutoff regulator, but such regulators do not preserve the gauge-invariance of the theory, which may introduce complications, which must be resolved.

**Acknowledgements**

We enjoyed several helpful discussions with S. Meuren, Profs. C.C. Gerry, A. Di Piazza, K.Z. Hatsagotsyan, C. Müller and Y.T. Li. QS and RG acknowledge the kind hospitality of their host institutions (Chinese Academy of Sciences, Beijing and MPIK Heidelberg) during their sabbatical leaves. This work has been supported by the NSF and the NSFC (#11128409).

**Appendix A**

A system is relativistically invariant if we are able to construct the Hamiltonian H, the total momentum P, and the boost operator K in such a way that they fulfill the three Poincare commutator relationships given by

$$[H,P] = 0 \tag{A1a}$$

$$[K,P] = -i\, H/c^2 \tag{A1b}$$

$$[K,H] = -i\, P \tag{A1c}$$

For the special case of a non-interacting system, the Hamiltonian is given by $H_0 = \int dp\, \omega(p)\, \hat{a}^\dagger(p)\hat{a}(p)$ and the total momentum operator by $P_0 = \int dp\, p\, \hat{a}^\dagger(p)\hat{a}(p)$. The generator for velocity boosts takes a more complicated form, $K_0 = i\int dp\, \{p/(2\omega(p))\, \hat{a}^\dagger(p)\hat{a}(p) + \omega(p)/c^2\, \partial \hat{a}^\dagger(p)/\partial p\, \hat{a}(p)\}$. This operator $K_0$ can also be represented in an equivalent form using the position operator $Z \equiv \int dz\, z\, \hat{a}^\dagger(z)\hat{a}(z)$, where $\hat{a}(z) \equiv (2\pi)^{-1/2} \int dp\, \hat{a}(p)\, \exp[ipz]$ is the Fourier transfrom of $\hat{a}(p)$. This position operator fullfills $[Z,P]=i$ and we can use it to express the (interaction-free) boost operator as $K_0 = -(ZH+HZ)/(2c^2)$. We note that the proof for $[K,H] = -i\, P$ requires us to assume that $\omega(p)\hat{a}^\dagger(p)\hat{a}(p)\big|_{-\infty}^{\infty} = 0$, i.e., that states with infinitely large momentum should not be excited.

Any operator, when viewed from another coordinate system which is shifted by a distance s, a time $\tau$, or a velocity v [associated with the rapidity parameter $\theta = \tanh(v/c)$], can be obtained from its original form by a unitary transformation. The quantum field operator $\hat{\varphi}(z,t)$ in Eq. (2.2) is constructed to have the remarkably simple transformation property

$$\exp[-iP_0 s]\, \hat{\varphi}(z,t)\, \exp[iP_0 s] = \hat{\varphi}(z-s,t) \tag{A2a}$$

$$\exp[iH_0 \tau]\, \hat{\varphi}(z,t)\, \exp[-iH_0 \tau] = \hat{\varphi}(z,t-\tau) \tag{A2b}$$

$$\exp[-iK_0 c\theta]\, \hat{\varphi}(z,t)\, \exp[iK_0 c\theta] = \hat{\varphi}(L_{-\theta}[z,t]) \tag{A2c}$$

where the linear operation $L_{-\theta}[z,t]$ describes the inverse of the usual Lorentz transformation,

which is defined as $L_\theta(a,b) \equiv (a\,\text{Cosh}\theta - b/c\,\text{Sinh}\theta,\ b\,\text{Cosh}\theta - a/c\,\text{Sinh}\theta)$.

For the interacting case of the $\hat{\varphi}(z)^N$ theory, we prove here that the new triplet, given by

$$H = H_0 + \lambda \int dz\, \hat{\varphi}(z)^N, \tag{A3a}$$

$$P = P_0 \tag{A3b}$$

$$K = K_0 + (\lambda/c^2) \int dz\, z\, \hat{\varphi}(z)^N \tag{A3c}$$

also fulfills the Poincare Lie algebra for any integer N except N=0. The proof is straightforward if we use the convenient transformation properties of the quantum field operator $\hat{\varphi}(z)^N$ described in Eqs. (A2).

The first Poincare relationship Eq. (A1a) can be proven as follows: $[H,P] = [H_0,P_0] + [\lambda \int dz\,\hat{\varphi}(z)^N, P_0] = [\lambda \int dz\,\hat{\varphi}(z)^N, P_0]$. To simplify the commutators further, we can use the general relationship between a general propagator $\exp[-iGa]$ and its associated generator G for arbitrary operators $O$ according to $\partial O/\partial a|_{a=0} = \partial\{\exp[-Ga]O\exp[Ga]\}/\partial a|_{a=0} = [O,G(a=0)]|_{a=0}$. We obtain

$$[\lambda \int dz\, \hat{\varphi}(z)^N, P_0] = \lim(s\to 0)\ \partial/\partial s\ \{\exp[-iP_0 s]\, \lambda \int dz\, \hat{\varphi}(z)^N\, \exp[iP_0 s]\}$$

$$= \lim(s\to 0)\ \lambda\, \partial/\partial s\, \int dz\, \hat{\varphi}(z+s)^N$$

$$= \lim(s\to 0)\ \lambda\, \partial/\partial s\, \int dz\, \hat{\varphi}(z)^N$$

$$= 0 \tag{A4}$$

Thus the first required relationship $[H,P]=0$ is fullfilled. The proof for the second relationship $[K,P]=-iH/c^2$ is very similar. We have $[K,P]= [K,P_0]=[K_0,P_0]+[(\lambda/c^2)\int dz\, z\, \hat{\varphi}(z)^N, P_0]$. For the second commutator we write:

$$[(\lambda/c^2)\int dz\, z\, \hat{\varphi}(z)^N, P_0] = \lim(s\to 0)\ i\, \partial/\partial a\ \{\exp[iP_0 s]\, (\lambda/c^2) \int dz\, z\, \hat{\varphi}(z)^N \exp[-iP_0 s]\}$$

$$= i\lim(s\to 0)\ (\lambda/c^2)\, \partial/\partial s\, \int dz\, z\, \hat{\varphi}(z+s)^N$$

$$= i \lim(s \to 0) \, (\lambda/c^2) \, \partial/\partial s \int dz \, (z-s) \, \hat{\varphi}(z)^N$$

$$= -i \lambda \int dz \, \hat{\varphi}(z)^N / c^2 \tag{A5}$$

Here it is important to note that Eq. (A5) is not valid for N=0 as the derivative of a constant operator $\int dz \, \hat{\varphi}(z)^0$ (even when unitarily transformed) is zero. As we can use for the first term $[K_0, P_0] = -iH_0/c^2$, we have therefore shown the validity of Eq. (A1b), i.e. $[K,P] = -iH/c^2$.

We now prove the third relationship Eq.(A1c), which is $[K,H] = -iP$. Again, inserting the expressions above from Eqs. (A3), we obtain the four terms $[K,H] = [K_0, H_0] + [K_0, \lambda \int dz \, \hat{\varphi}(z)^N] + [(\lambda/c^2) \int dz \, z \, \hat{\varphi}(z)^N, H_0] + [(\lambda/c^2) \int dz \, z \, \hat{\varphi}(z)^N, \lambda \int dz \, \hat{\varphi}(z)^N]$. The first term is $-iP_0$ while the fourth term vanishes. The remaining second and third terms are

$$[K_0, \lambda \int dz \, \hat{\varphi}(z,t=0)^N] = \lim(\theta \to 0) \, i \, \partial/\partial(c\theta) \, \{\exp[-iK_0 c\theta] \, \lambda \int dz \, \hat{\varphi}(z,t=0)^N \exp[iK_0 c\theta]\} \tag{A6}$$

If we use $\exp[-iK_0 c\theta] \, \hat{\varphi}(z,t=0) \exp[iK_0 c\theta] = \hat{\varphi}(z \cosh\theta, z/c \sinh\theta)$, the commutator becomes

$$\begin{aligned}[K_0, \lambda \int dz \, \hat{\varphi}(z,t=0)^N] &= i \lim(\theta \to 0) \, \partial/\partial(c\theta) \, \{\lambda \int dz \, \hat{\varphi}(z \cosh\theta, z/c \sinh\theta)^N\} \\
&= i \lim(\theta \to 0) \, c^{-1} \{\lambda \int dz \, [\, z \sinh\theta \, \partial/\partial z \, \hat{\varphi}(z \cosh\theta, t=z/c \sinh\theta)^N \\
&\quad + [z/c \cosh\theta \, \partial/\partial t \, \hat{\varphi}(z\cosh\theta, t=z/c \sinh\theta)^N \} \\
&= i \lambda/c^2 \int dz \, z \, \partial/\partial t \, \hat{\varphi}(z,t)^N \big|_{t=0} \end{aligned} \tag{A7}$$

For the third commutator, we apply the same reasoning

$$\begin{aligned}[(\lambda/c^2) \int dz \, z \, \hat{\varphi}(z)^N, H_0] &= \lim(t \to 0) \, -i \, \partial/\partial t \, \{\exp[-iH_0 t] \, (\lambda/c^2) \int dz \, z \, \hat{\varphi}(z)^N \exp[iH_0 t]\} \\
&= \lim(t \to 0) \, -i \, \partial/\partial t \, \{(\lambda/c^2) \int dz \, z \, \hat{\varphi}(z,t)^N\} \\
&= -i(\lambda/c^2) \int dz \, z \, \partial/\partial t \, \hat{\varphi}(z,t)^N \big|_{t=0} \end{aligned} \tag{A8}$$

As the sum of the two terms in Eqs. (A7) and (A8) cancel, we have the final result $[K,H] =$

30        10/10/12

$-iP_0 = -iP$.

Finally, we point out that in any relativistic theory describing particles of mass m, the spectrum of the Hamiltonian operator H must satisfy two conditions: the spectrum should be non-negative, and it should contain 0 as a non-degenerate eigenvalue. These conditions follow from the existence of the no-particle (vacuum) state |vac⟩ which carries the trivial 1-dimensional irreducible representation of the Poincare group, i.e., H |vac⟩ = P |vac⟩ = K |vac⟩ = 0. If any Hamiltonian H is a legitimate Hamiltonian satisfying these conditions, then for any non-zero constant N the operator H′=H+N does not satisfy these conditions and is not a legitimate Hamiltonian. First, note that any eigenvector of H is automatically an eigenvector of H′. Therefore, the spectrum of H′ can be obtained from the spectrum of H by simply shifting it by the constant N. If N were positive, then the spectrum of H′ would violate the condition that 0 should be an eigenvalue, and if N were negative, then the spectrum violates the condition that the spectrum should be non-negative.

Even though it might seem only as a trivial energy shift, we have not been able to construct the corresponding velocity boost generator K which would make the Hamiltonian $H_0 + N$, with N a constant, fulfill the Poincare relationships, which is strictly required for relativistic invariance. While this infinite constant could be important for phenomena in systems with finite extension (such as the Casimir effect. e.g.), a system with a boundary would not be translationally invariant and we presently do not know if it is possible to generalize the Poincare relationships to account for the effective interactions that are usually approximated mathematically by boundary conditions. In Appendix A we review this invariance based on the Poincare relationships in more detail.

**Appendix B**

The two terms in the interaction V which modify the energy of a single-particle state $|p\rangle$ at second order in $\lambda$ are proportional to $\hat{a}^\dagger \hat{a}^\dagger \hat{a}^\dagger \hat{a}$, corresponding an intermediate 3-particle state, and $\hat{a}^\dagger \hat{a} \hat{a}^\dagger \hat{a}^\dagger$, where now the intermediate state will be a 5-particle state. We will define the two corresponding second-order energy corrections as $E^{(2)}_3$ and $E^{(2)}_5$. For the term $E^{(2)}_3$, the matrix element to be calculated is

$$\langle p_1 p_2 p_3 | V | p \rangle = \langle p_1 p_2 p_3 | (2\pi) [(4\pi)^{-1/2} c]^4 \lambda \int dk_1 dk_2 dk_3 dk_4 \, \delta(-k_1-k_2-k_3+k_4)$$

$$\times [\omega(k_1)\omega(k_2)\omega(k_3)\omega(k_4)]^{-1/2} \, 4 \, \hat{a}^\dagger(k_1) \hat{a}^\dagger(k_2) \hat{a}^\dagger(k_3) \hat{a}(k_4) \hat{a}^\dagger(p) |0\rangle$$

$$= (2\pi) [(4\pi)^{-1/2} c]^4 \lambda \, \langle 0| \hat{a}(p_1) \hat{a}(p_2) \hat{a}(p_3) / f_{123} \int dk_1 dk_2 dk_3 dk_4 \, \delta(-k_1-k_2-k_3+k_4)$$

$$[\omega(k_1)\omega(k_2)\omega(k_3)\omega(k_4)]^{-1/2} \, 4 \, \hat{a}^\dagger(k_1) \hat{a}^\dagger(k_2) \hat{a}^\dagger(k_3) \delta(k_4 - p) |0\rangle$$

$$= (2\pi) [(4\pi)^{-1/2} c]^4 \lambda \, \langle 0| \hat{a}(p_1) \hat{a}(p_2) \hat{a}(p_3) / f_{123} \int dk_1 dk_2 dk_3 \, \delta(-k_1-k_2-k_3+p)$$

$$[\omega(k_1)\omega(k_2)\omega(k_3)\omega(p)]^{-1/2} \, 4 \, \hat{a}^\dagger(k_1) \hat{a}^\dagger(k_2) \hat{a}^\dagger(k_3) |0\rangle$$

$$= 3! \, 4 \, (2\pi) [(4\pi)^{-1/2} c]^4 \lambda \, f_{123}^{-1} \, \delta(-p_1-p_2-p_3+p) [\omega(p_1)\omega(p_2)\omega(p_3)\omega(p)]^{-1/2} \langle 0|0\rangle$$

(B.1)

The factor of 3! in the final line arises from the number of ways in which the $\hat{a}(p_i)$ can contract on the $\hat{a}^\dagger(k_i)$, and $f_{123} \equiv f(p_1, p_2, p_3)$ is the combinatorical factor which arises from the operators $\hat{a}^\dagger(p_i)$ acting on the bare vacuum state $|0\rangle$; $f_{123}$ is equal to 1 if all three momenta differ, it is equal to $\sqrt{2}$ if two momenta are equal to each other (but not all three), and it is equal to $\sqrt{6}$ if all three momenta are identical.

This matrix element is to be integrated over all possible 3-particle intermediate states $|p_1 p_2 p_3\rangle$, $\int d\beta = \int dp_1 \int dp_2 \int dp_3$, where the integration range is restricted to $p_1 \leq p_2 \leq p_3$ to avoid double counting. The integration range can be extended to include all momenta if we also divide by the symmetry factor $M_{123} \equiv M(p_1, p_2, p_3)$, which counts the number of times the triplet $p_1, p_2, p_3$ appears in the integration (or summation, in the case of a finite box). This symmetry factor works out to be $M_{123} = 3! / f_{123}^2$. It might appear that a combinatorical factor such as $f_{123}$ is irrelevant inside an integral, since it requires two or more momenta to be identical for it to be different from 1, which is of measure 0. However, if the integral is divergent, a measure 0 contribution to an integral may yield a finite contribution. While the integral $E^{(2)}_3$ is finite, the integral $E^{(2)}_5$ diverges because it contains a term equal to the vacuum's energy correction. This vacuum energy correction is to be subtracted from all states in the Hilbert space, and in order to do this correctly finite corrections to these divergent integrals must be treated very carefully. The energy correction corresponding to 3-particle states is therefore

$$E^{(2)}_3 = \delta(0)^{-1} \iiint dp_1 dp_2 dp_3 \, M_{123}^{-1} \, | \, 3! \, 4 \, (2\pi) \, [(4\pi)^{-1/2} c]^4 \lambda \, f_{123}^{-1} \, \delta(-p_1-p_2-p_3+p)$$
$$\times [\omega(p_1)\omega(p_2)\omega(p_3)\omega(p)]^{-1/2} \, |^2 \, / \, (\omega_p - (\omega(p_1)+\omega(p_2)+\omega(p_3)))$$
$$= 3/(2\pi^2) \, c^8 \lambda^2 \iint dp_1 dp_2 \, [\omega(p_1)\omega(p_2)\omega(p-p_1-p_2)\omega(p)]^{-1} \, / \, (\omega_p - \omega(p_1) - \omega(p_2) - \omega(p-p_1-p_2))$$

(B.2)

The factor of $\delta(-p_1-p_2-p_3+p)$ is squared; one delta function removes an integral, and the other cancels the factor of $\delta(0)^{-1}$ which comes from the normalization of the state $|p\rangle$, yielding a finite correction to the energy. The computation of $E^{(2)}_5$ proceeds similarly. The matrix element to be computed is

$$\langle p_1 p_2 p_3 p_4 p_5 | V | p \rangle = \langle p_1 p_2 p_3 p_4 p_5 | (2\pi) [(4\pi)^{-1/2} c]^4 \lambda \int dk_1 \int dk_2 \int dk_3 \int dk_4 \, \delta(k_1+k_2+k_3+k_4)$$
$$[\omega(k_1)\omega(k_2)\omega(k_3)\omega(k_4)]^{-1/2} \, \hat{a}^\dagger(k_1) \hat{a}^\dagger(k_2) \hat{a}^\dagger(k_3) \hat{a}^\dagger(k_4) \, \hat{a}^\dagger(p) |0\rangle$$
$$= \langle 0 | \hat{a}(p_1)\hat{a}(p_2)\hat{a}(p_3)\hat{a}(p_4)\hat{a}(p_5) f_{12345}^{-1} | (2\pi) [(4\pi)^{-1/2} c]^4 \lambda \int dk_1 \int dk_2 \int dk_3 \int dk_4 \, \delta(k_1+k_2+k_3+k_4)$$
$$[\omega(k_1)\omega(k_2)\omega(k_3)\omega(k_4)]^{-1/2} \, \hat{a}^\dagger(k_1)\hat{a}^\dagger(k_2)\hat{a}^\dagger(k_3)\hat{a}^\dagger(k_4) \, \hat{a}^\dagger(p) |0\rangle$$
$$= (2\pi) [(4\pi)^{-1/2} c]^4 \lambda \, f_{12345}^{-1} \, ( \, \delta(p_5-p) \, 4! \, \delta(p_1+p_2+p_3+p_4) [\omega(p_1)\omega(p_2)\omega(p_3)\omega(p_4)]^{-1/2}$$
$$+ \delta(p_4-p) \, 4! \, \delta(p_1+p_2+p_3+p_5) [\omega(p_1)\omega(p_2)\omega(p_3)\omega(p_5)]^{-1/2}$$
$$+ \delta(p_3-p) \, 4! \, \delta(p_1+p_2+p_4+p_5) [\omega(p_1)\omega(p_2)\omega(p_4)\omega(p_5)]^{-1/2}$$

$$+ \delta(p_2-p) \, 4! \, \delta(p_1+p_3+p_4+p_5) \, [\omega(p_1)\omega(p_3)\omega(p_4)\omega(p_5)]^{-1/2}$$

$$+ \delta(p_1-p) \, 4! \, \delta(p_2+p_3+p_4+p_5) \, [\omega(p_2)\omega(p_3)\omega(p_4)\omega(p_5)]^{-1/2}) \quad \text{(B.3)}$$

Each of these 5 terms comes from a contraction on one of the momenta $p_1$, $p_2$, $p_3$, $p_4$, or $p_5$ of the intermediate particles contracting on $\hat{a}^\dagger(p)|0\rangle$, while the factor of 4! comes from all of the different (and equivalent) ways in which the remaining operators can contract on each other. The 5-particle combinatorical factor $f_{12345} \equiv f(p_1,p_2,p_3,p_4,p_5)$ is defined in a similar way as it was for the 3-particle case. This matrix element must once again be integrated over all possible intermediate states $\int d\beta = \int dp_1 \int dp_2 \int dp_3 \int dp_4 \int dp_5$, where the integration range is restricted to $p_1 \leq p_2 \leq p_3 \leq p_4 \leq p_5$ to avoid double counting. Here again the restriction on the integration range can be lifted if we divide by a symmetry factor which avoids double counting, which is $M_{12345} \equiv M(p_1,p_2,p_3,p_4,p_5) = 5!/f_{12345}^2$. The energy correction is therefore

$$E^{(2)}_5 = \delta(0)^{-1} \iiiiint dp_1 dp_2 dp_3 dp_4 dp_5 \, M_{12345}^{-1} \, \Big| (2\pi) [(4\pi)^{-1/2} c]^4 \lambda \, f_{12345}^{-1}$$

$$\times \delta(p_5-p) \, 4! \, \delta(p_1+p_2+p_3+p_4) \, [\omega(p_1)\omega(p_2)\omega(p_3)\omega(p_4)]^{-1/2} + \text{4 permutations} \Big|^2 /$$

$$(\omega_p - (\omega_1+\omega_2+\omega_3+\omega_4+\omega_5))$$

$$= \delta(0)^{-1} \iiiiint dp_1 dp_2 dp_3 dp_4 dp_5 \, f_{12345}^2/5! \, (1/(8\pi) \, c^4 \lambda \, f_{12345}^{-1} \, 4!)^2$$

$$\times \Big| \delta(p_5-p) \, \delta(p_1+p_2+p_3+p_4) \, [\omega(p_1)\omega(p_2)\omega(p_3)\omega(p_4)]^{-1/2} + \text{4 permutations} \Big|^2 /$$

$$(\omega_p - (\omega_1+\omega_2+\omega_3+\omega_4+\omega_5)) \quad \text{(B.4)}$$

When this 5-term polynomial is squared, there will be 25 terms; 5 of these correspond to each term itself squared, and 20 will be cross-terms. Fortunately, the 5 individual terms squared are all equal to each other, and the cross-terms are likewise all equal to one another. An individual term squared gives

$$= \delta(0) \, 3/(40\pi^2) \, c^8 \lambda^2 \iiiint dp_1 dp_2 dp_3 \, [\omega(p_1)\omega(p_2)\omega(p_3)\omega(-p_1-p_2-p_3)]^{-1} / (-\omega_1-\omega_2-\omega_3-\omega_4)$$

When calculating this term, each of the two delta functions $\delta(p_5-p)$ and $\delta(p_1+p_2+p_3+p_4)$ removes one integration, but since these delta functions are squared there is a remaining factor of $\delta(0)^2$, which makes the entire term proportional to $\delta(0)$. There are 5 such terms, and when Eq. (B.5) is multiplied by 5, it is found to be exactly the second order correction to the vacuum energy, $E^{(2)}_{vac}$. $E^{(2)}_{vac}$ can be calculated by a very similar perturbative calculation as presented above, but beginning with the state $|0\rangle$ instead of $|p\rangle$. Since this vacuum energy is to be removed from all states in the theory by subtracting a counterterm from the Hamiltonian, Eq (B.5) is not to be intepreted as a contribution to the single particle mass.

The cross-terms that come from the square in Eq. (B.4) are all equal by permutations, and are of the form

$$= \delta(0)^{-1} \iiiint\!\!\int dp_1 dp_2 dp_3 dp_4 dp_5 \, f_{12345}{}^2/5! \, (1/(8\pi)) \, c^4 \lambda \, f_{12345}{}^{-1} \, 4!)^2$$

$$\times \delta(p_5-p)\delta(p_1+p_2+p_3+p_4)\delta(p_4-p)\,\delta(p_1+p_2+p_3+p_5)\,[\omega(p_1)\omega(p_2)\omega(p_3)]^{-1}[\omega(p_4)\,\omega(p_5)]^{-1/2}/$$

$$(\omega_p-(\omega_1+\omega_2+\omega_3+\omega_4+\omega_5))$$

$$= \delta(0)^{-1} \, 3/(40\pi^2) \, c^8\lambda^2 \iiint dp_1 dp_2 dp_3 \, \delta(p_1+p_2+p_3+p)\,\delta(p_1+p_2+p_3+p)$$

$$\times [\omega(p_1)\omega(p_2)\omega(p_3)]^{-1}[\omega(p)\omega(p)]^{-1/2}/(-\omega_1-\omega_2-\omega_3-\omega_p)$$

$$= 3/(40\pi^2) \, c^8\lambda^2 \iint dp_1 dp_2 \, [\omega(p_1)\omega(p_2)\omega(-p-p_1-p_2)\omega(p)]^{-1}/(-\omega(p_1)-\omega(p_2)-\omega(-p-p_1-p_2)-\omega(p))$$

(B.6)

This term is finite, as the structure of the delta functions leaves only one delta function with an argument of 0, and this canels the $\delta(0)^{-1}$ which came from the normalization of $|p\rangle$. There are 20 cross-terms such as Eq. (B.6), all of which differ only by a permutation of integration variables. The total correction to the energy, if the vacuum term Eq. (B.5) is subtracted, is equal to Eq. (B.2) plus 20 times the cross-term Eq. (B.6), which is

$$E^{(2)}(p) = E^{(2)}_3 + E^{(2)}_5 - E^{(2)}_{vac}$$

$$= 3/(2\pi^2) \, c^8 \lambda^2 \iint dp_1 dp_2 \, [\omega(p_1)\omega(p_2)\omega(p-p_1-p_2)\omega(p)]^{-1} / (\omega_p - \omega(p_1) - \omega(p_2) - \omega(p-p_1-p_2))$$

$$+ 3/(2\pi^2) \, c^8 \lambda^2 \iint dp_1 dp_2 \, [\omega(p_1)\omega(p_2)\omega(-p-p_1-p_2)\omega(p)]^{-1} / (-\omega(p_1) - \omega(p_2) - \omega(-p-p_1-p_2) - \omega(p))$$

(B.7)

The correction to the mass can be evaluated by evaluating $E^{(2)}(p=0)$ numerically, which for a bare mass m=1 gives a mass correction of $-\lambda^2 \times 0.75$. This result agrees very well with the data obtained from the Feynman graphs in Section 5.1.2.

**Appendix C**

As the scattering-matrix approach presented here is based on computing the matrix elements of the scattering operator for the vacuum $|0\rangle$ as well as single-particle state $|q\rangle$ with monetum q, it is advantageous to have a direct relationship between the matrix elements of $|0\rangle$ and $|q\rangle$, such that we only need to calculate the easier elements for the vacuum state to know their values for single particle states. Below we proof the useful relationships Eqs. (C1) and (C3). In first-order perturbation theory, the calculation of the scattering-matrix elements requires us to compute the expectation values of powers of the operators $\hat{\varphi}(z,t)$ and $D(\hat{\varphi}) \equiv (\partial_{ct}\hat{\varphi}(z,t))^2 - (\partial_z\hat{\varphi}(z,t))^2$.

We will first prove the general relationship

$$\langle q_1|\hat{\varphi}^N|q_2\rangle = \langle 0|\hat{\varphi}^N|0\rangle \delta(q_1-q_2) + (N-1)N \langle 0|\hat{\varphi}^N|0\rangle Y(q_1)Y^*(q_2) \qquad (C1)$$

where the c-number $Y(q_j) \equiv [\hat{a}(q_1), \hat{\varphi}(z,t)] = c\,(4\pi)^{-1/2}\,\omega(q_1)^{-1/2}\,\exp[-iq_1 z+i\omega(q_1)t]$. Note that for the special case $q_1=q_2$ this reduces to

$$\langle q|\hat{\varphi}^N|q\rangle = \langle 0|\hat{\varphi}^N|0\rangle\,\delta(0) + (N-1)N\,c^2\,[4\pi\omega(q_1)]^{-1}\,\langle 0|\hat{\varphi}^N|0\rangle \qquad (C2)$$

There are two steps involved in the derivation. First we commute $\hat{a}(q_1)$ in $\langle 0|\hat{a}(q_1)\hat{\varphi}^N|q_2\rangle$ to the right, obtaining $\langle 0|\hat{a}(q_1)\hat{\varphi}^N|q_2\rangle = \langle 0|\hat{\varphi}^N \hat{a}(q_1)|q_2\rangle + \langle 0|[\hat{a}(q_1), \hat{\varphi}^N]|q_2\rangle$. The first term simplifies to $\langle 0|\hat{\varphi}^N|0\rangle\,\delta(q_1-q_2)$. As the commutator $[\hat{a}(q_1), \hat{\varphi}]=Y(q_1)$ is a c-number and therefore commutes with $\hat{\varphi}^N$, we can simplify the term $[\hat{a}(q_1), \hat{\varphi}^N] = N\,Y(q_1)\,\hat{\varphi}^{N-1}$. This can easily be seen from the solution to the recursive relationship $[\hat{a}(q_1), \hat{\varphi}^N] = \hat{\varphi}[\hat{a}(q_1), \hat{\varphi}^{N-1}] + [\hat{a}(q_1), \hat{\varphi}]\hat{\varphi}^{N-1}$. So our two terms simplify to $\langle 0|\hat{a}(q_1)\hat{\varphi}^N|q_2\rangle = \langle 0|\hat{\varphi}^N|0\rangle\,\delta(q_1-q_2) + N\,Y(q_1)\langle 0|\hat{\varphi}^{N-1}|q_2\rangle$.

As a second step we commute $\hat{a}^\dagger(q_2)$ in $\langle 0|\hat{\varphi}^{N-1}|q_2\rangle = \langle 0|\hat{\varphi}^{N-1}\hat{a}^\dagger(q_2)|0\rangle$ to the leftmost

position, leading to $\langle 0|\hat{\varphi}^{N-1}\hat{a}^{\dagger}(q_2)|0\rangle = \langle 0|[\hat{\varphi}^{N-1}, \hat{a}^{\dagger}(q_2)]|0\rangle + \langle 0|\hat{a}^{\dagger}(q_2)\hat{\varphi}^{N-1}|0\rangle$, where the second term vanishes. Following the same sort of recursion relationship argument as above, we find that $[\hat{\varphi}^{N-1}, \hat{a}^{\dagger}(q_2)] = (N-2)\, Y^*(q_2)\, \hat{\varphi}^{N-1}$, such that the final correction term becomes $N\, Y(q_1)\langle 0|\hat{\varphi}^{N-1}|q_2\rangle = (N-1)(N-2)\, Y(q_1)\, Y^*(q_2)\, \langle 0|\hat{\varphi}^{N-2}|0\rangle$ and the proof of Eq. (C1) is complete.

A second relationship between the vacuum and single-particle state expectation value is given by

$$\langle q_1|D(\hat{\varphi})^N|q_2\rangle = \langle 0|D(\hat{\varphi})^N|0\rangle\, \delta(q_1-q_2) +$$

$$+ N\, 2(T(q_1)T^*(q_2) - Z(q_1)Z^*(q_2))\, \langle 0|D(\hat{\varphi})^{N-1}|0\rangle$$

$$+ 4N(N-1)\, \langle 0|[T(q_1)\partial_{ct}\hat{\varphi} - Z(q_1)\partial_z\hat{\varphi}][T^*(q_2)\partial_{ct}\hat{\varphi} - Z^*(q_2)\partial_z\hat{\varphi}]D(\hat{\varphi})^{N-2}|0\rangle \quad (C3)$$

Here we have defined the operators $\hat{A}(q_j) \equiv [\hat{a}(q_j), D(\hat{\varphi})] = 2[T(q_j)\partial_{ct}\hat{\varphi} - Z(q_j)\partial_z\hat{\varphi}]$ and the two c-numbers $Z(q_j) \equiv [\hat{a}(q_j), \partial_z\hat{\varphi}] = \partial_z Y(q_j)$ and $T(q_j) \equiv [\hat{a}(q_j), \partial_z\hat{\varphi}] = \partial_{ct} Y(q_j)$. For the interesting special case $q_1 = q_2 \equiv q$ this simplifies to

$$\langle q|D(\hat{\varphi})^N|q\rangle = \langle 0|D(\hat{\varphi})^N|0\rangle\, \delta(0) +$$

$$+ N\, (2\pi)^{-1}\, m^2 c^4/\omega(q)\, \langle 0|D(\hat{\varphi})^{N-1}|0\rangle$$

$$+ N(N-1)\, \pi^{-1}\, \omega(q)\, \langle 0|(\partial_{ct}\hat{\varphi} + cq/\omega(q)\, \partial_z\hat{\varphi})^2\, D(\hat{\varphi})^{N-2}|0\rangle \quad (C4)$$

To prove this relationship, we follow a similar two-step approach as above. First, we again commute $\hat{a}(q_1)$ in $\langle 0|\hat{a}(q_1)D(\hat{\varphi})^N|q_2\rangle$ to the right, obtaining $\langle 0|\hat{a}(q_1)D(\hat{\varphi})^N|q_2\rangle = \langle 0|D(\hat{\varphi})^N\hat{a}(q_1)|q_2\rangle + \langle 0|[\hat{a}(q_1), D(\hat{\varphi})^N]|q_2\rangle$. The first term simplifies to $\langle 0|D(\hat{\varphi})^N|0\rangle\, \delta(q_1-q_2)$. As the commutator $[\hat{a}(q_1), D(\hat{\varphi})] = \hat{A}(q_1)$ is an operator that commutes with $D(\hat{\varphi})$, $[\hat{A}(q_j), D(\hat{\varphi})] = 0$, we can simplify the term $[\hat{a}(q_1), D(\hat{\varphi})^N] = N\, \hat{A}(q_1)\, D(\hat{\varphi})^{N-1}$. This can be easily seen from the solution to the recursive relationship $[\hat{a}(q_1), D(\hat{\varphi})^N] = D(\hat{\varphi})[\hat{a}(q_1), D(\hat{\varphi})^{N-1}] + [\hat{a}(q_1), D(\hat{\varphi})] D(\hat{\varphi})^{N-1}$. So the second terms simplifies to $\langle 0| [\hat{a}(q_1), D(\hat{\varphi})^N]|q_2\rangle = N \langle 0|\hat{A}(q_1) D(\hat{\varphi})^{N-1} |q_2\rangle$.

The second step is to commute $\hat{a}^\dagger(q_2)$ in $N \langle 0|\hat{A}(q_1) D(\hat{\varphi})^{N-1} |q_2\rangle = N \langle 0|\hat{A}(q_1) D(\hat{\varphi})^{N-1} \hat{a}^\dagger(q_2)|0\rangle$ to the leftmost position, leading to

$$N \langle 0| \hat{A}(q_1) D(\hat{\varphi})^{N-1} \hat{a}^\dagger(q_2)|0\rangle$$
$$= N \langle 0|\hat{A}(q_1)\hat{a}^\dagger(q_2)D(\hat{\varphi})^{N-1}|0\rangle + N \langle 0|\hat{A}(q_1) [D(\hat{\varphi})^{N-1}, \hat{a}^\dagger(q_2)]|0\rangle \quad (C5)$$

Using the commutator $[D(\hat{\varphi})^{N-1}, \hat{a}^\dagger(q_2)] = (N-1) \hat{A}^*(q_2) D(\hat{\varphi})^{N-2}$, this simplifies to

$$N \langle 0| \hat{A}(q_1) D(\hat{\varphi})^{N-1} \hat{a}^\dagger(q_2)|0\rangle$$
$$= N \langle 0|\hat{A}(q_1)\hat{a}^\dagger(q_2)D(\hat{\varphi})^{N-1}|0\rangle + N(N-1) \langle 0|\hat{A}(q_1) \hat{A}^*(q_2) D(\hat{\varphi})^{N-2}|0\rangle$$
$$= N \langle 0|[\hat{A}(q_1),\hat{a}^\dagger(q_2)] D(\hat{\varphi})^{N-1}|0\rangle + N(N-1) \langle 0|\hat{A}(q_1) \hat{A}^*(q_2) D(\hat{\varphi})^{N-2}|0\rangle \quad (C6)$$

The commutator $[\hat{A}(q_1),\hat{a}^\dagger(q_2)] = 2[(T(q_1)\partial_{ct}\hat{\varphi} - Z(q_1)\partial_z\hat{\varphi}), \hat{a}^\dagger(q_2)]$ simplifies to $2(T(q_1)T^*(q_2) - Z(q_1)Z^*(q_2))$. If we use the definitions of $T(q_j)$ and $Z(q_j)$, this commutator simplifies to $(2\pi)^{-1} (\omega(q_1)^{1/2} \omega(q_2)^{1/2} + c^2 q_1 q_2 \omega(q_1)^{-1/2} \omega(q_2)^{-1/2}) \exp[-iq_1 z + i\omega(q_1)t] \exp[iq_2 z - i\omega(q_2)t]$. Furthermore, the term $\hat{A}(q_1) \hat{A}^*(q_2)$ is a short-hand notation for the operator $\hat{A}(q_1) \hat{A}^*(q_2) = 4[T(q_1)\partial_{ct}\hat{\varphi} - Z(q_1)\partial_z\hat{\varphi}][T^*(q_2)\partial_{ct}\hat{\varphi} - Z^*(q_2)\partial_z\hat{\varphi}]$. We therefore obtain the final result as indicated by Eq. (C3):

$$\langle q_1|D(\hat{\varphi})^N|q_2\rangle = \langle 0|D(\hat{\varphi})^N|0\rangle \delta(q_1-q_2) + N\, 2(T(q_1)T^*(q_2) - Z(q_1)Z^*(q_2)) \langle 0|D(\hat{\varphi})^{N-1}|0\rangle$$
$$+ 4N(N-1) \langle 0|[T(q_1)\partial_{ct}\hat{\varphi} - Z(q_1)\partial_z\hat{\varphi}][T^*(q_2)\partial_{ct}\hat{\varphi} - Z^*(q_2)\partial_z\hat{\varphi}]D(\hat{\varphi})^{N-2}|0\rangle \quad (C7)$$

This expression simplifies considerably if we assume $q_1=q_2\equiv q$. In this case, the second expression $T(q_1)T^*(q_2) - Z(q_1)Z^*(q_2)=(4\pi)^{-1} (\omega(q)-c^2q^2/\omega(q))$ reduces to the factor $(4\pi)^{-1} m^2c^4/\omega(q)$. Furthermore, the product of the two operators $[T(q_1)\partial_{ct}\hat{\varphi}- Z(q_1)\partial_z\hat{\varphi}][T(q_2)\partial_{ct}\hat{\varphi}- Z(q_2)\partial_z\hat{\varphi}]$ reduces to $(4\pi)^{-1} \omega(q)[\partial_{ct}\hat{\varphi}+cq/\omega(q)\,\partial_z\hat{\varphi}]^2$. If we insert these expressions into Eq. (C7) we obtain Eq. (C4).

**References**

[1]  I. Bialynicki-Birula and Z. Bialynickia-Birula, "Quantum electrodynamics" (Pergamon Press, Oxford, 1975).

[2]  C.C. Gerry and P.L. Knight, "Introductory quantum optics" (Cambridge University Press, Oxford, 2004).

[3]  J. Collins, "Renormalization" (Cambridge University Press, Cambridge, 2003).

[4]  B. Delamotte, Am. J. Phys. 72, 170 (2004).

[5]  N.N. Bogoliubov and D.V. Shirkov, "The theory of quantized fields" (Interscience, New York, 1959).

[6]  S.S. Schweber, "An introduction to relativistic quantum field theory" (Harper & Row, New York, 1962).

[7]  P. Roman, "Introduction to quantum field theory" (John Wiley& Sons, New York, 1969).

[8]  C. Itzykson and J. Zuber, "Quantum field theory" (McGraw-Hill, New York, 1980).

[9]  L.H. Ryder, "Quantum Field Theory" (Cambridge University Press, 1985).

[10] S. Weinberg, "The quantum theory of fields", Vol. 1 (Cambridge University Press, Cambridge, England, 1995).

[11] M.E. Peskin and D.V. Schroeder, "An introduction to quantum field theory" (Westview Press, Boulder, 1995).

[12] M. Srednicki, "Quantum field theory" (Cambridge University Press, Cambridge, 2007).

[13] F. Mandl and G. Shaw, "Quantum field theory" (Wiley, Chichester, 2006).

[14] K. Takayanagi, Phys. Rev. A 44, 59 (1991).

[15] E.V. Stefanovich, Ann. Phys. 292, 139 (2001).

[16] E.V. Stefanovich, "Renormalization and dressing in quantum field theory", hep-th/0503076 (2005).

[17] E.V. Stefanovich, "Relativistic quantum dynamics: A non-traditional perspective on space, time, particles, fields, and action-at-a-distance" (Mountain View, 2004), available at arXiv:physics/0504062v13 [physics.gen-ph].

[18] G.V. Efimov, Int. J. Mod. Phys. A4, 4977 (1989).

[19] J. Glimm and A. Jaffe, Phys. Rev. 176, 1945 (1968).

[20] A.M. Jaffe and R.T. Powers, Commun. Math. Phys. 7, 218 (1968).

[21] J. Glimm and A. Jaffe, Ann. Math. 91, 362 (1970).

[22] R.E. Wagner, S. Acosta, S.A. Glasgow, Q. Su and R. Grobe, J. Phys. A (in press).

[23]  M. Kaku, "Quantum field theory" (Oxford University Press, Oxford, 1993).

[24]  A. Messiah, "Quantum mechanics", Vol. II (Noth-Holland, Amsterdam, 1966).

[25]  T. Cheng, E.R. Gospodarczyk, Q. Su and R. Grobe, Ann. Phys. 325, 265 (2010).

[26]  I. Montvay and G. Münster, "Quantum fields on a lattice" (Cambridge University Press, Cambridge, 1994).